\documentclass[pra,aps,twocolumn,longbibliography,10pt%
]{revtex4-1}

\usepackage{graphicx}
\usepackage{amsmath, amsxtra, amssymb, mathtools, isomath, nccmath, xspace, xcolor}

\usepackage[utf8]{inputenc}
\usepackage[T1]{fontenc}  

\usepackage{times}
\usepackage{ wasysym }
\usepackage{xfrac} 
\usepackage{bm}	%

\usepackage{setspace, changepage} 
\usepackage{etoolbox}

\usepackage[normalem]{ulem}

\definecolor{myColor}{rgb}{0.02,0.12,0.3}
\definecolor{myciteColor}{rgb}{0.39,0.7,0.89}
\usepackage[%
colorlinks=true,%
citecolor=myColor,%
linkcolor=myColor,%
urlcolor=myColor%
]{hyperref}
\usepackage[all]{hypcap} 

\renewcommand{\k}[2]{\ensuremath{^{#1#2}\textrm{K}}}

\newcommand{\mfp}[0]{\ensuremath{\ell_{\textrm{mfp}}}\xspace}

\newcommand{\braces}[1]{\ensuremath{\left( #1 \right)}\xspace}
\newcommand{\ket}[1]{\ensuremath{\left| #1 \right\rangle}\xspace}

\newcommand{\avg}[1]{\left\langle #1 \right\rangle}

\newcommand{\drvTD}[3]{\left(\frac{\partial #1}{\partial #2}\right)_{#3} }
\newcommand{\pvec}[1]{\begin{pmatrix} #1 \end{pmatrix}}
\newcommand{\eqtext}[1]{\ensuremath{\stackrel{\text{#1}}{=}}}

\newcommand{\kB}[0]{\ensuremath{k_{\textrm B}}}
\newcommand{\vn}[0]{\ensuremath{v_{\textrm n}}}
\newcommand{\vneq}[0]{\ensuremath{v_{\textrm n,0}}}
\newcommand{\vs}[0]{\ensuremath{v_{\textrm s}}}
\newcommand{\vseq}[0]{\ensuremath{v_{\textrm s,0}}}
\newcommand{\nn}[0]{\ensuremath{n_{\textrm n}}}
\newcommand{\nneq}[0]{\ensuremath{n_{\textrm n,0}}}
\newcommand{\ns}[0]{\ensuremath{n_{\textrm s}}}
\newcommand{\nBEC}[0]{\ensuremath{n_{\textrm{BEC}}}}
\newcommand{\nseq}[0]{\ensuremath{n_{\textrm s,0}}}
\newcommand{\Tc}[0]{\ensuremath{T_{\textrm c}}}
\newcommand{\kc}[0]{\ensuremath{k_{\textrm c}}}

\newcommand{\cc}[0]{\ensuremath{c_{\rm cl}}}
\newcommand{\var}[2][-0.07]{\ensuremath{\delta\kern#1em#2}}
\newcommand\umu{\ensuremath{\textrm{µ}}}
\newcommand{\kunit}[0]{\ensuremath{\umu\textrm{m}^{\text{--}1}}}
\newcommand{\s}[0]{\ensuremath{_{\rm s}}}
\newcommand{\n}[0]{\ensuremath{_{\rm n}}}

\def\nobreakbefore{%
  \relax\ifvmode\else
    \ifhmode
      \ifdim\lastskip > 0pt\relax
        \unskip\nobreakspace
      \else 
        \nobreakspace
      \fi
    \fi
  \fi
}
\let\oldcite\cite
\renewcommand\cite{\nobreakbefore\oldcite}

\begin{document}
\title{
First and second sound in a compressible 3D Bose fluid
}
\author{
Timon A.\ Hilker,$^{1,{\color{myColor}\ast}}$
Lena H.\ Dogra,$^{1}$
Christoph Eigen,$^{1}$
Jake A.\ P.\ Glidden,$^{1}$
Robert P.\ Smith,$^{2}$ and
Zoran Hadzibabic$^{1}$
}
\affiliation{
\vspace{1.5mm}
$^1$ Cavendish Laboratory, University of Cambridge, J.\ J.\ Thomson Avenue, Cambridge CB3 0HE, United Kingdom\\
$^2$ Clarendon Laboratory, University of Oxford, Parks Road, Oxford OX1 3PU, United Kingdom
}

\begin{abstract}	
The two-fluid model is fundamental for the description of superfluidity. 
In the nearly-incompressible-liquid regime, it successfully describes first and second sound, corresponding, respectively, to density and entropy waves, in both liquid helium and unitary Fermi gases.
Here, we study the two sounds in the opposite regime of a highly compressible fluid, using an ultracold $^{39}$K Bose gas in a three-dimensional box trap.
We excite the longest-wavelength mode of our homogeneous gas,
and observe two distinct resonant oscillations below the critical temperature, of which only one persists above it. 
In a microscopic mode-structure analysis, we find agreement with the hydrodynamic theory, where first and second sound involve density oscillations dominated by, respectively, thermal and condensed atoms.
Varying the interaction strength, we explore the crossover from hydrodynamic to collisionless behavior in a normal gas.
\end{abstract}

\maketitle

One of the hallmarks of superfluidity is the existence of two distinct sound modes with the same wavelength, corresponding to two different sound speeds. 
This remarkable property is the key prediction of the hydrodynamic two-fluid model, which was conceptualized by Tisza\cite{Tisza:1938} and London\cite{London:1938}, and established by Landau using quantum hydrodynamics\cite{Landau:1941,Landau:1941b}. 
In this model, the two fluids are the superfluid and the normal component of a system below its critical temperature $\Tc$.
Originally inspired by superfluid $^{4}$He, 
Landau's theory successfully predicts properties of this strongly interacting, essentially incompressible fluid. 
More recently, two sound modes have been observed in ultracold atomic Fermi gases near unitarity\cite{Sidorenkov:2013,Hoffmann:2020,Zwierlein:2020}, which are also nearly incompressible.

\begin{figure}[!b]
\centering
\includegraphics[width=0.95\columnwidth]{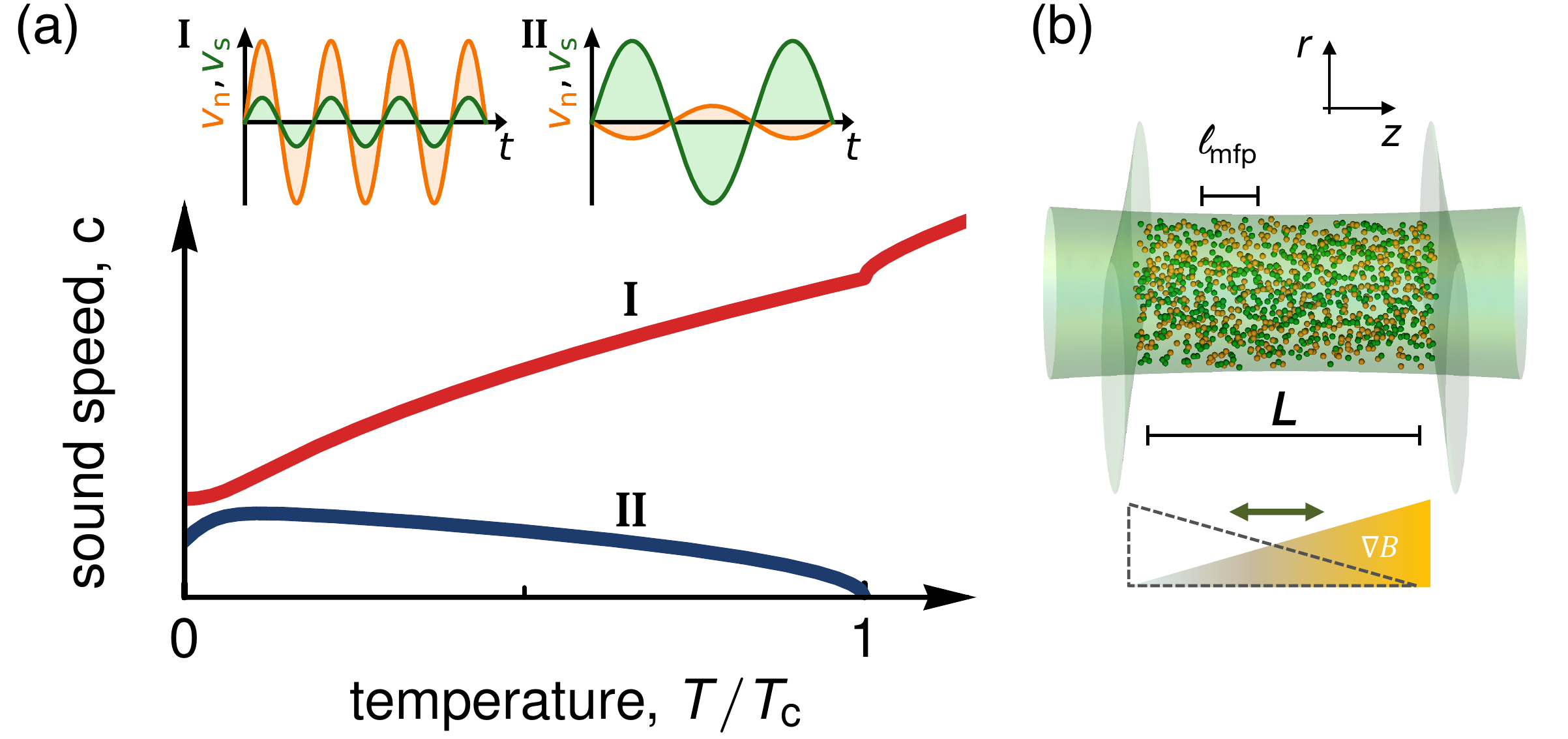}
\caption{
First and second sound in dilute Bose gases.
(a) Mode structure and speeds of sound in a weakly interacting Bose gas based on the two-fluid model.
Both modes involve motion of both fluids ($\vs, \vn$), but for $\kB T\gg g \ns$, where $g$ is the interaction strength, the first (second) sound is mainly an oscillation of the normal (superfluid) component.
(b) In a box-trapped gas of \k39 atoms, we excite the $k_L=\pi/L$ mode of both sounds by sinusoidal forcing using a magnetic gradient $\nabla B$. To attain hydrodynamic conditions, we reduce the mean free path $\mfp$ below the box length $L$ by tuning $g$.
}
\label{fig_1}
\end{figure}

Ultracold Bose gases provide a versatile platform to test the same general framework for highly compressible superfluids, including the hybridization of the two modes\cite{Pitaevskii2016} [see \hyperref[fig_1]{Fig.\,1(a)}]. 
However, a challenge in these systems is reaching hydrodynamic conditions for the normal fluid, which requires the collisional mean free path $\mfp$ to be significantly shorter than the excitation wavelength.
In harmonically trapped gases, following first studies of collisionless excitations\cite{jin96coll,mewe96coll,Andrews:1997}, a pioneering study~\cite{Stamper-Kurn:1998} revealed the analogues of first and second sound in two collective modes with frequencies between the hydrodynamic and collisionless predictions.
Further studies have explored the effects of interactions on the first sound above $\Tc$ \cite{Leduc:2002,Buggle2005} and on the second sound and related thermodynamics below it\cite{Meppelink2009,Ville2018,Fritsch2018}.
Recently, simultaneous observation of first and second sound has been used to characterize the superfluid transition in two dimensions (2D)\cite{Christodoulou2020}.

Here, we realize the compressible-fluid regime of the two-fluid model in the textbook setting of a 3D homogeneous Bose gas, using tuneable interactions between $^{39}$K atoms to attain hydrodynamic conditions. 
Below $\Tc$, we observe both first and second sound, with speeds in agreement with Landau's theory.
Using momentum-resolved measurements, which give access to the motion of the spatially overlapping superfluid and normal components, we reveal the structure of the two sound modes.
For gases above $\Tc$, where only the first sound remains, we also investigate the effects of viscous damping by reducing the interaction strength and crossing from the hydrodynamic to the collisionless regime.

\begin{figure*}
\centering
\includegraphics[width=0.99\textwidth]{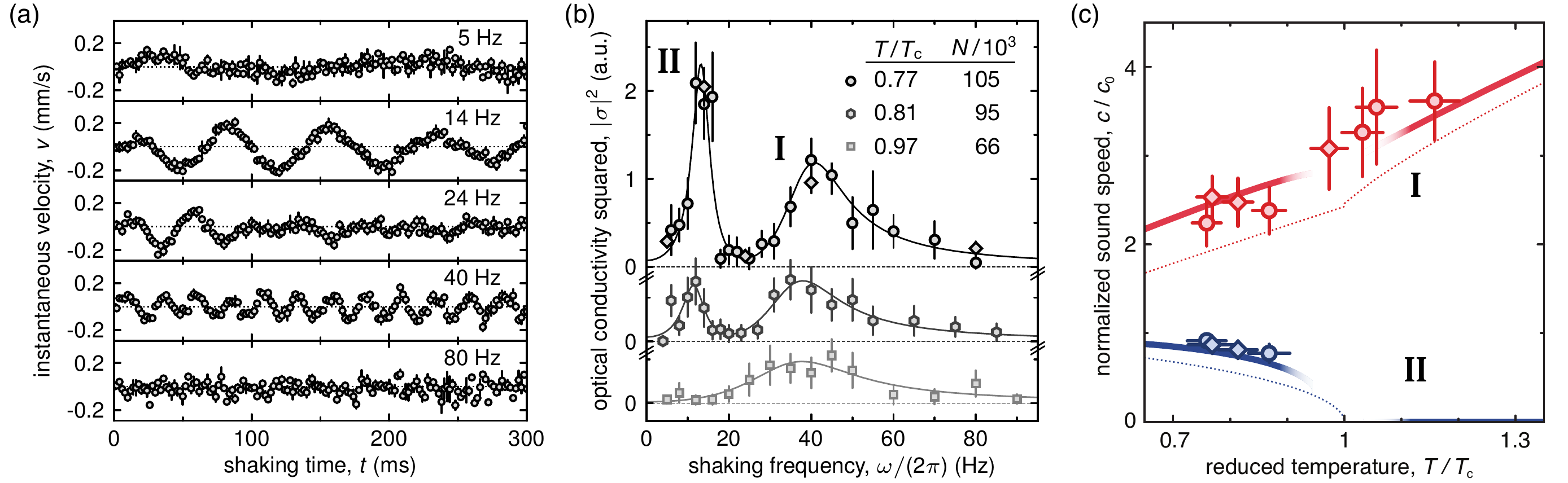}
\caption{
Observation of first and second sound.
(a) Center-of-mass velocity $v$ versus shaking time $t$ at several frequencies $\omega/(2\pi)$ for a cloud with $N=105(3)\times 10^{3}$ atoms and $T=97(3)\,$nK (measured at $t=200\,$ms).
(b) $|\sigma(\omega)|^2$, where $|\sigma(\omega)|$ is proportional to the amplitude of the total current $N v(t)$. The two peaks correspond to the two sound modes. We show spectra for different $N$ at approximately the same $T$.
The solid lines are fits using the sum of two resonances\cite{SI}. The diamonds correspond to the data shown in (a). 
(c)~First (red) and second (blue) sound speeds, normalized by the Bogoliubov speed $c_0(N)$, versus $T/T_{\rm c}$.
The diamonds correspond to data shown in (b).
The solid lines are the fit-free predictions of the two-fluid model [\hyperref[eq_cH]{Eq.\,(\ref*{eq_cH})}] at fixed $T=97$\,nK and varying $N$, with $\ns$ calculated using Popov mean-field theory.
Near $\Tc$ mean-field theory is not valid, which we represent by fading of the lines.
The dotted lines show $c_{\rm I}$ of a non-interacting gas and $c_{\rm II}$ using the ideal-gas condensate density for \ns.
}
\label{fig_2}
\end{figure*}


In the hydrodynamic two-fluid model, the superfluid and normal components are characterized by their densities ($\ns,\nn$) and velocities ($\vs,\vn$). 
In the nearly incompressible liquid helium, the first sound is an in-phase oscillation of the two coupled components, while the second sound is an out-of-phase oscillation that corresponds to a pure temperature or entropy per particle wave.
On the other hand, in a strongly compressible gas the two components largely decouple.
As illustrated in \hyperref[fig_1]{Fig.\,1(a)}, in this regime the first and second sound, respectively, are predominantly $\vn$ and $\vs$ modes\cite{Lee1959,Pitaevskii2016}.

Generally (see \cite{SI} for details), one can write an eigenvalue equation~\cite{Landau:1941,Nozieres:1990} for the sound speed $c$ in the basis of total density $n$ and entropy per particle $\tilde{s}$:
\begin{align}
b^2\pvec{
J^2/K^2                 &                  J  \\[8pt]
J  &  1
}
 \pvec{\frac{\delta n}{n} \\[8pt] \frac{\delta\tilde{s}}{\tilde{s}}}
&=
c^2
\pvec{1 & 0\\[8pt] 
0 & \tfrac{\nn}{\ns}
}
\pvec{\frac{\delta n}{n} \\[8pt] \frac{\delta\tilde{s}}{\tilde{s}}} ,
\label{eqs_couplingGeneral}
\end{align} 
where $K^2 \equiv (1-c_V/c_P)$\footnote{The parameter $K^2=\varepsilon/(1+\varepsilon)$ is directly related to the Landau--Placzek ratio $\varepsilon=c_P/c_V -1 $~\cite{Landau:1934}}\nocite{Landau:1934}, $J \equiv -\frac{n}{\tilde{s}}\drvTD{\tilde{s}}{n}{T,V}$, and $b^2 \equiv nT\tilde{s}^2/(mc_V)$, with heat capacities $c_{V,P}$ per volume $V$, temperature $T$, and atom mass $m$.
For an incompressible gas, $K \rightarrow 0$ and \mbox{$J\sim1$} for any $T<\Tc$, so the two modes are oscillations of $n$ and $\tilde{s}$~\cite{Pitaevskii2016}. 
In the opposite, ideal-gas limit~\footnote{We understand the ideal gas as the limit $a\to 0^+$. The strictly non-interacting gas has $c_\mathrm{II}=0$ and is not a superfluid.}, $K,J\rightarrow1$, the $n$ and $\tilde{s}$ modes maximally hybridize; here the eigenmodes correspond to motion of either the normal or the superfluid component.
In general, the crossover between the two regimes is primarily controlled by the value of $K$, which can vary between zero and one.

In a weakly interacting Bose gas, where thermodynamic quantities can be calculated from first principles,
$K$ changes smoothly from zero to almost one as the temperature is varied from zero to $T_\mathrm{c}$\cite{Pitaevskii2016}.
Here we explore the compressible regime, $\kB T\gg gn$, where $g = 4 \pi \hbar^2 a /m$ and $a$ is the \mbox{$s$-wave} scattering length.
In Hartree--Fock mean-field theory, $K = 1-g n^2/[2\chi(z)\nn\kB T]$ and the sound speeds are\cite{Griffin:1997,SI}
\begin{align}
c_{\,\rm I}^{(\rm HF)}= \sqrt{\chi(z)\frac{\kB T}{m}+2\frac{g\nn}{m}}\,,\quad c_{\,\rm II}^{(\rm HF)} = \sqrt{\frac{g \ns}{m}}\,\label{eq_cH},
\end{align}
where $\chi(z)=5g_{5/2}(z)/(3g_{3/2}(z))$, $g_{\alpha}(z)$ are polylogarithms, and $z=e^{\mu^*/(\kB T)}$ is evaluated using $\mu^*=-g\ns$ below $\Tc$.
While $c_{\,\rm II}^{(\rm HF)}$ depends strongly on interactions, $c_{\,\rm I}^{(\rm HF)}$ is to leading order set by just the temperature.


\begin{figure*}
\centering
\includegraphics[width=0.98\textwidth]{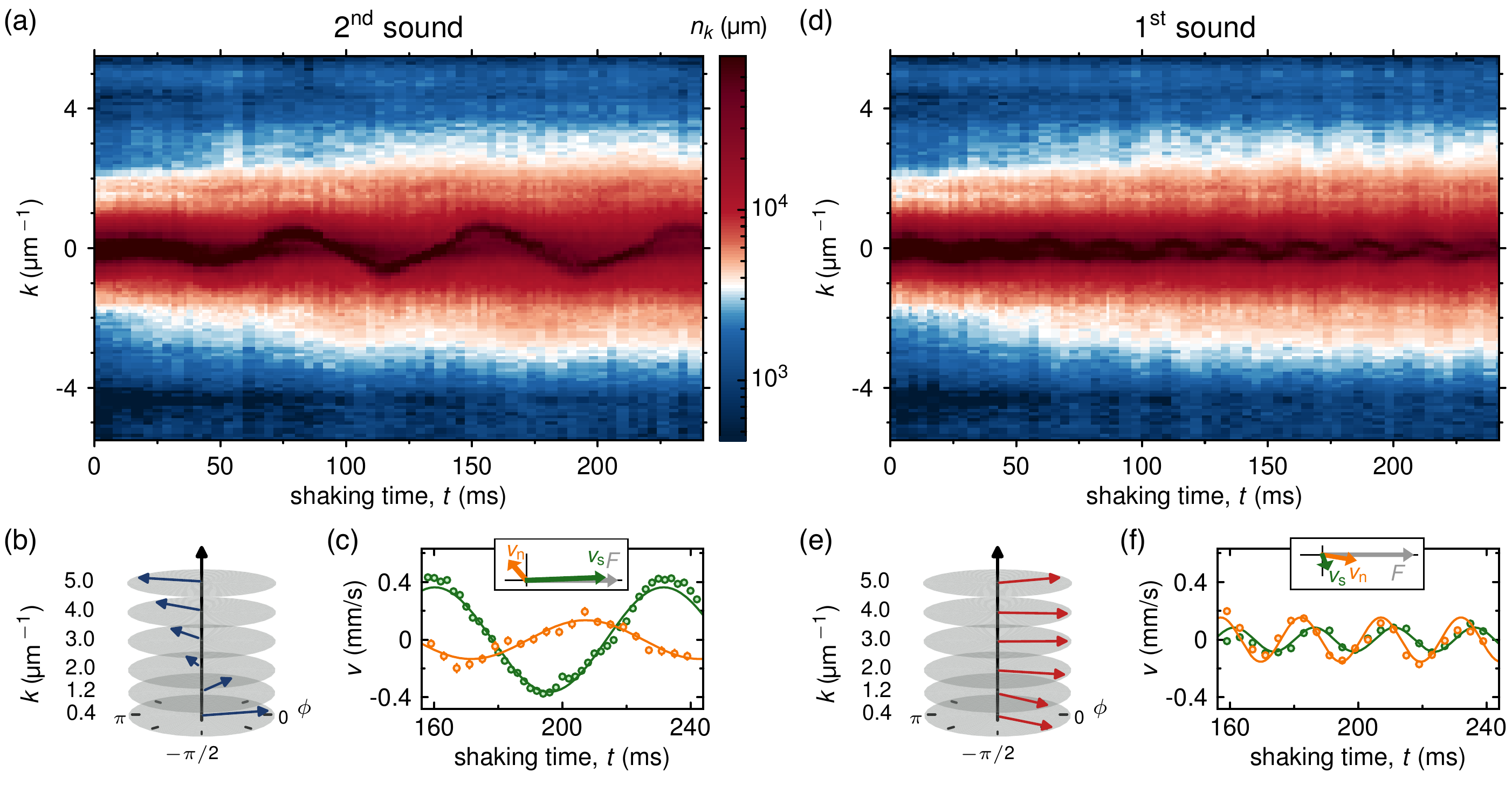}
\caption{
Microscopic structure of the two sound modes. Here $T/T_{\rm c}=0.77(3)$, corresponding to a superfluid fraction of 43(3)\%. 
(a) Evolution of $n_k$ on the second-sound resonance at 14\,Hz [see \hyperref[fig_2]{Fig.\,2(b)}]. The motion of the superfluid at low $k$ is visually clear, and from the wings of the distribution one can also extract the normal-gas velocity $\vn$. 
(b) $k$-resolved phase $\phi$ (w.r.t.\ the phase of the drive) of the $n_k(t)$ oscillations.
 Here the low- and high-$k$ oscillations are almost completely out of phase, as predicted for second sound.
(c) Extracted $\vn$ (orange) and $\vs$ (green); $\vs$ is larger and in phase with the drive (see phasor in the inset) as expected for a compressible superfluid.
(d-f) Analogous analysis to (a-c), now for the first-sound resonance at 40 Hz; here $\phi$ is close to $0$ for all $k$ and $\vn$ is larger than $\vs$. 
}
\label{fig_3}
\end{figure*}


Our experiments start with a partially condensed Bose gas of \k39 atoms in the lowest hyperfine state, confined in a cylindrical box trap \cite{Gaunt:2013,Eigen:2016,Navon:2021} of length $L=70(2)\,\umu$m and radius $R=9.2(5)\,\umu$m. To create hydrodynamic conditions we tune $a$ to a relatively high 480(20)\,$a_0$ using the magnetic Feshbach resonance at $402.7$\,G~\cite{Fletcher:2017}. This enhances three-body losses and heating, but within $200$\,ms the gas reaches a trap-depth--limited $T= 97(3)$\,nK (see \cite{SI}), at which point the atom number is $N =105(3)\times10^3$, 
corresponding to $T/\Tc=0.77(3)$ \footnote{Throughout the paper, we evaluate $\protect \ensuremath
  {T_{\protect \textrm c}}$ based on ideal-gas theory. The expected
  interaction-induced shift of $\protect \ensuremath {T_{\protect \textrm c}}$
  is on the level of one percent in our system (see \cite{SI}).}, 
\mbox{$\mfp=(8\pi n a^2)^{-1}=0.15(1) L$}, $K=0.75(5)$, $J=1.20(4)$, and $b=3.2(2)\,$mm\,s$^{\text{--}1}$.

After tuning $a$, we start exciting the lowest sound mode(s), with wavevector $k_{L}=\pi/L$, using a spatially uniform force of magnitude $F=F_0\sin(\omega t)$, with \mbox{$F_0 L/k_\text{B}=7.7\,\text{nK}$}, generated by an axial magnetic gradient~\cite{Navon2016}.
After a variable time $t$, we turn off both $F$ and the trap, and measure the axial 1D momentum distribution $n_k(t)$ using a time-of-flight expansion of $30\text{--}45$\,ms.
The cloud's center-of-mass (CoM) velocity $v(t)=\hbar \langle k \rangle/m$, shown in \hyperref[fig_2]{Fig.\,2(a)} for various $\omega$, gives the current density due to both components, $j = nv = \ns\vs+\nn\vn$. This provides a model-free description of the system response, which we characterize in \mbox{(quasi-)steady} state ($t\gtrsim 200\,$ms) using the analogue of a complex optical conductivity $\sigma \propto j/F$\cite{Anderson2017}. 

In \hyperref[fig_2]{Fig.\,2(b)} (top curve) we plot the full spectrum $|\sigma(\omega)|^2$ at $T/\Tc=0.77$, which reveals two well resolved peaks corresponding to the two sound modes.
In our homogeneous system, the resonance frequencies $\omega_{\rm I}$ and $\omega_{\rm II}$ directly give the speeds $c_{\rm I,II}=\omega_{\rm I,II}/k_L $.
To study sound modes at higher $T/\Tc$, we lower $N$ at approximately fixed $T$\cite{SI}.
We see that the response at $\omega_{\rm II}$ weakens and vanishes, while the response at $\omega_{\rm I}$ persists [lower curves of \hyperref[fig_2]{Fig.\,2(b)}].
 
In \hyperref[fig_2]{Fig.\,2(c)} we summarize our measurements of $c_{\rm I,II}$ for various $T/\Tc$ below and above 1. Here, we normalize the speeds by the Bogoliubov speed $c_0=[gN/(\pi R^2Lm)]^{1/2}$. We find good agreement with the predictions of \hyperref[eq_cH]{Eq.\,(\ref*{eq_cH})} without any free parameters (solid lines), with $\ns$ calculated within the mean-field Popov approximation \cite{SI}.
For reference, we also show the ideal-gas prediction for $c_{\rm I}$~
\footnote{The interaction shift of $c_{\rm I}$ is noticable in our system even though the interactions are only just strong enough to reach the hydrodynamic regime. This can be contrasted to air at room temperature, which is very hydrodynamic [$\lambda/\mfp\sim$ $na^2\,\lambda$ $\approx10^7$ using $\lambda\approx 1\,$m and $\mfp\approx100\,$nm], but almost a perfect ideal gas [$gn/(\kB T)\sim na\lambda_T^2\approx 10^{-5}$ using an effective $a$ based on $\mfp$] due to the much higher temperature.} and the prediction for $c_{\rm II}$ assuming that $\ns$ is given by the non-interacting condensate density (dotted lines). 


We next explore the $k$-space structure of the two sound modes, focusing on our $T/\Tc=0.77$ dataset, for which the superfluid fraction is 43(3)\%. 
\hyperref[fig_3]{Figures\,3(a,d)} show the time evolution of the axial $n_k$ distribution for shaking near the second- and first-sound resonance (at 14~Hz and 40~Hz respectively); for details of the experimental procedure see~\cite{SI}.
In both cases the motion of the low-$k$ superfluid component is visually more clear, and it is more pronounced for the second sound.

We quantitatively analyze these data in two ways. 
First, in \hyperref[fig_3]{Fig.\,3(b,e)}, we simply look at the phase $\phi$ of the occupation-number oscillation at each $k$, fitting a sinusoid to $n_k(t)-n_{-k}(t)$ and defining $\phi$ relative to the drive. For second sound, $\phi$ wraps from 0 to $\pi$ with increasing $k$. On the other hand, for first sound, $\phi$ is close to zero for all $k$.
Second, we disentangle the motion of normal and superfluid components, which both contribute at low $k$. 
We deduce $\vn(t)$ by looking only at high $k$ ($>1.7\,\kunit$), assuming that the hydrodynamic motion of the normal component corresponds to a simple displacement of its whole momentum distribution, and then calculate $\vs(t)$ from $\vn(t)$, the total current, and the superfluid fraction~\cite{SI}. 
These results are shown in \hyperref[fig_3]{Fig.\,3(c,f)}.
Note that $\vs$ at the first-sound resonance (40~Hz) could be slightly affected by the proximity to the second-sound resonance for \mbox{$k=3k_L$}.

These observations demonstrate all the key features of the two-fluid theory for a highly compressible gas.
Additional information is contained in the damping of the modes, seen in their nonzero widths [see \hyperref[fig_2]{Fig.\,2(b)}]; zero hydrodynamic damping would require the collision rate to be infinite, to ensure instantaneous local equilibration, which is not the case even in gases with infinite scattering length\cite{Patel2020,Bohlen2020,Wang2021}. 

For the second sound we deduce an amplitude-damping coefficient $\gamma_{\rm II}=2\pi\times 2.7(4)\,$Hz, with no clear $N$-dependence. This is compatible with the Landau--Khalatnikov hydrodynamic prediction~\cite{Hohenberg1965,Nikuni2001,SI}~$\gamma_{\rm II}\approx 2\pi\times 2.2\,$Hz. It also coincides with the Landau-damping prediction~\cite{Griffin2009} $3\pi a k_L k_{\rm B}T/(8\hbar)=2\pi\times2.7(2)\,$Hz, but this is likely fortuitous since that theory assumes a collisionless normal component. 
Note that in the experiment additional broadening may arise due to nonlinear effects~\cite{Zhang:2021} and the temporal variation of the gas density caused by losses; in the future it would be interesting to study the damping of the second sound further.

For the first sound, we systematically explore the crossover from hydrodynamic to collisionless regime in gases above $\Tc$, where hydrodynamic behavior relies entirely on scattering. The parameter separating the two regimes is the Knudsen number, which in our case is given by ${\mfp/L}$.


\begin{figure}[!t]
\centering
\includegraphics[width=1.0\columnwidth]{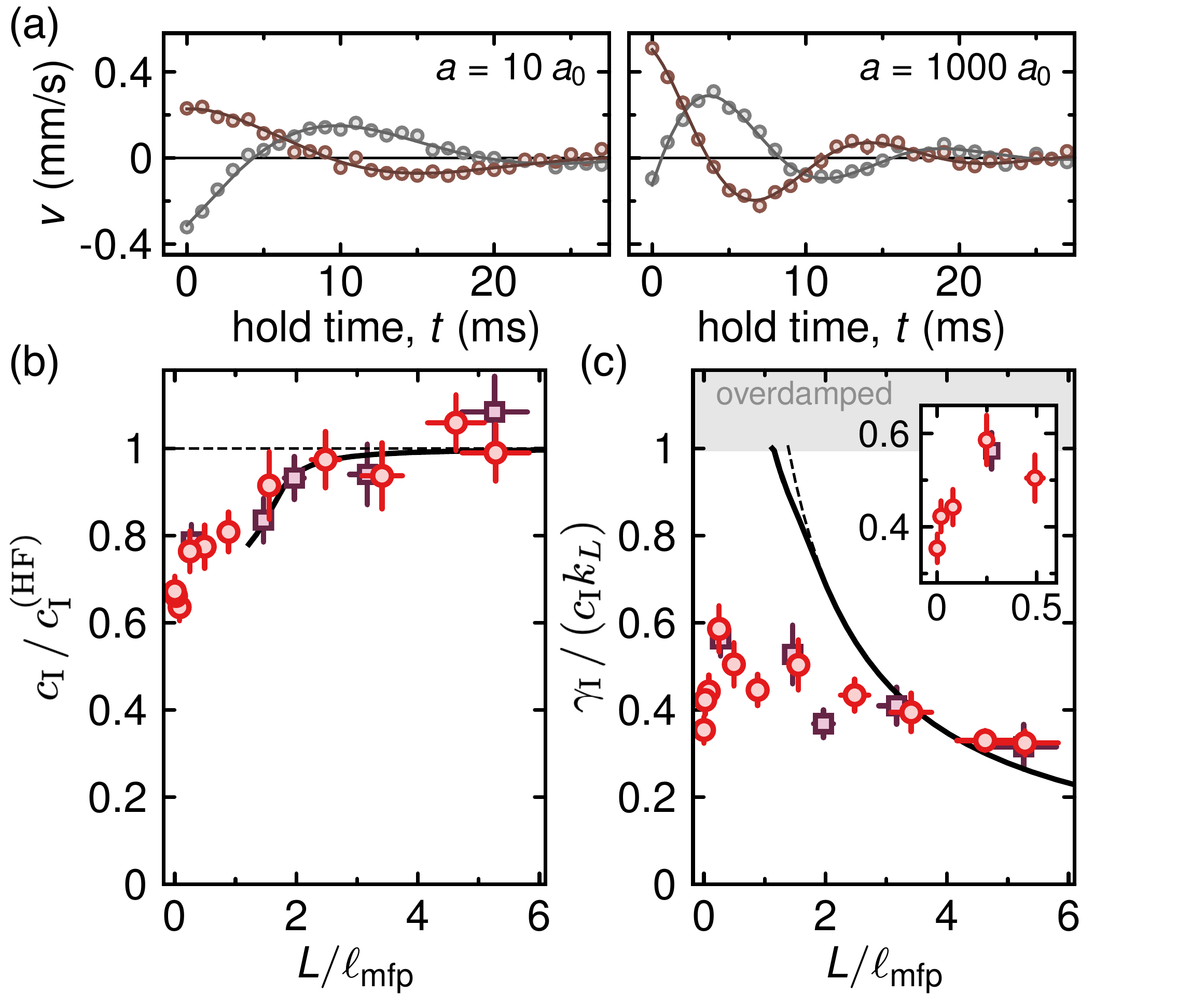}
\caption{
From collisionless dynamics to hydrodynamic sound in a normal gas. (a) For scattering lengths from $a=10\,a_0$ to 1000\,$a_0$, we measure the CoM velocity $v$ for a free oscillation after shaking the cloud for 3 (grey circles) and 3.25 (brown circles) periods at $55\,$Hz. 
(b) Measured speed of sound $c_{\rm I}$ normalized to the speed $c_{\rm I}^{\rm (HF)}$ predicted for dissipationless hydrodynamics. (c) Damping per period. In both (b) and (c), the data for $L=70\,\umu$m (purple squares) coalesce with the data taken using an additional box geometry ($L=50\,\umu$m, red circles), when plotted against the inverse Knudsen number $L/\mfp\sim na^2/k_L$.
The dashed lines show theoretical predictions to linear order in $\mfp/L$, while the solid lines show the results of the full hydrodynamic calculation (see text and \cite{SI}).
The latter captures the observed drop of the normalized sound speed below unity.  
The relative damping also agrees with this fit-free theory at $L/\mfp>3$, while for $L/\mfp\to0$ it decreases but remains nonzero (see zoom-in inset).
}
\label{fig_4}
\end{figure}

We prepare gases at $T\approx1.3\,\Tc$ and vary $\mfp/L$ both by tuning $a$ and using two different box lengths ($50\,\umu$m and $70\,\umu$m).
We initiate a CoM velocity oscillation by shaking at $55$\,Hz with $F_0 = \kB \times 0.55\,$nK$\,\umu$m$^{\text{--}1}$, then stop the drive and observe decaying free oscillations [see \hyperref[fig_4]{Fig.\,4(a)}]. 
We extract the sound speed $c_{\rm I}$ and damping $\gamma_{\rm I}$ by fitting $v(t)$ with the harmonic-oscillator form \mbox{$v\propto\cos(\omega_{\rm I} t+\phi)e^{-\gamma_{\rm I} t}$}, with $\omega_{\rm I}=\sqrt{(c_{\rm I}k_L)^2-\gamma_{\rm I}^2}$.
We normalize $c_{\rm I}$ by its prediction in the hydrodynamic limit $c_{\rm I}^{\rm (HF)}$ [\hyperref[eq_cH]{Eq.~(\ref*{eq_cH})}], and $\gamma_{\rm I}$ by the angular frequency $c_{\rm I}k_L$.

Theoretically, to linear order in $\mfp/L$, the sound speed retains its hydrodynamic value, but the nonzero heat conductivity $\kappa$ and viscosity $\eta$ result in a nonzero $\gamma_{\rm I}$. 
The predicted damping, or equivalently the diffusivity $D_{\rm I}=2\gamma_{\rm I}/k^2$, is given by the Stokes--Kirchhoff relation $D_{\rm SK}= [4\eta/(3mn)+\kappa(c_V^{-1}-c_P^{-1})]$\cite{Landau1987}. 
In kinetic gas theory $\eta/(mn)\sim\kappa/c_P\sim \mfp c_{\rm I}$. For a weakly interacting gas, $c_{\rm I}\sim \sqrt{\kB T/m}$, so $D_{\rm I}\sim(\mfp/\lambda_T)\frac{\hbar}{m}$, where $\lambda_T$ is the thermal wavelength, and one can write $\gamma_{\rm I}/(c_{\rm I}k)= r k\mfp$, where the dimensionless $r$ depends on the degeneracy; for our $T/\Tc$ we get $r\approx0.44$~\cite{SI,Nikuni1997}.
This theory, without free parameters, is shown in \hyperref[fig_4]{Fig.\,4(b,c)} as the dashed line and it agrees with our data for $L/\mfp\gtrsim3$. 

For $L/\mfp<3$ the measured $c_{\rm I}$ decreases, while $\gamma_{\rm I}$ keeps growing albeit less than predicted. 
Using the classical sound equations, we calculate the effects of $\eta$ and $\kappa$ on $c_{\rm I}$ and $\gamma_{\rm I}$ beyond linear order (see solid lines in \hyperref[fig_4]{Fig.\,4(b,c)}, and \cite{SI,Temkin1981} for details).
The higher-order effects depend on the Prandtl number, $\text{Pr}$, which measures the relative weights of momentum and thermal diffusivities. 
For our system $\text{Pr}\approx 2/3$~\cite{SI,Nikuni1997}, and the prediction is for $c_{\rm I}$ to decrease, in agreement with our data.
The predicted damping is also reduced, but not significantly.

In the collisionless regime $L/\mfp\to0$, where the hydrodynamic theory does not apply, the measured damping is finite and the oscillations are still well described by an exponential damping. Note that, unlike for the shape oscillations in harmonic traps\cite{Pitaevskii2016,Buggle2005}, a nonzero damping is expected even at $a=0$ due to the dispersion of momentum modes.

Finally, we note the distinction between $c_{\rm I}$ and the phase velocity $\omega/k$.
While $c_{\rm I}$ is a material property, the effect of viscous damping on $\omega/k$ depends on the boundary conditions; for our fixed-wavelength case $\omega/k$ decreases with damping, whereas it increases in the fixed-frequency case~\cite{SI,Temkin1981}.


In conclusion, we observed both first and second sound in a 3D ultracold Bose gas that is sufficiently strongly interacting to be hydrodynamic, but is still highly compressible. 
We found that Landau's two-fluid theory captures all the essential features of this system, with the first and second sound mode, respectively, predominantly featuring oscillations of the normal and the superfluid component.
By tuning interactions, we also studied the breakdown of hydrodynamicity. The experimental access to both microscopic and hydrodynamic properties offers an excellent opportunity for further studies of Bose fluids. In particular, it would be interesting to explore lower temperatures ($k_{\rm B}T\approx g\ns$) where an avoided crossing of the two sound modes is expected\cite{Pitaevskii2016}. Below this crossing the incompressible limit is approached and the Bogoliubov mode becomes the first sound, while the nature of the second mode is still subject of theoretical investigation\cite{Seifie2019}.


We thank Panagiotis Christodoulou, Ludwig Mathey, Sandro Stringari, and Martin Zwierlein for discussions. This work was supported by EPSRC [Grants No.~EP/N011759/1 and No.~EP/P009565/1], ERC (QBox), and a QuantERA grant (NAQUAS, EPSRC Grant No.~EP/R043396/1). T.~A.~H. acknowledges support from the EU Marie Sk\l{}odowska-Curie program [Grant No.~MSCA-IF- 2018 840081]. C.~E. acknowledges support from Jesus College (Cambridge). R.~P.~S. acknowledges support from the Royal Society. Z.~H. acknowledges support from the Royal Society Wolfson Fellowship.
\vspace{-1.em}



\begingroup
\renewcommand{\addcontentsline}[3]{}

%
\endgroup

\clearpage

{\centering\bf  \large Supplemental Material for:\\
\vspace{0.1cm}
\centering
First and second sound in a compressible 3D Bose fluid\\}

\setcounter{figure}{0} 
\setcounter{equation}{0} 

\renewcommand\theequation{S\arabic{equation}} 
\renewcommand\thefigure{S\arabic{figure}}

\tableofcontents
\vspace{1em}
Supporting material for Eqs.~(1) and (2) of the main paper is provided in Section~\ref{sec_si_theory}.
The methods used for Fig.~2 and Fig.~3 are described in Section~\ref{sec_exp_protocol}.
The theory presented in Fig.~4 is derived in Section~\ref{sec_si_damping}.

\section{Dissipationless two-fluid model:\\ from incompressible to compressible fluids}
\label{sec_si_theory} 

\begin{figure}[!t]
\centering
\includegraphics[width=0.99\columnwidth]{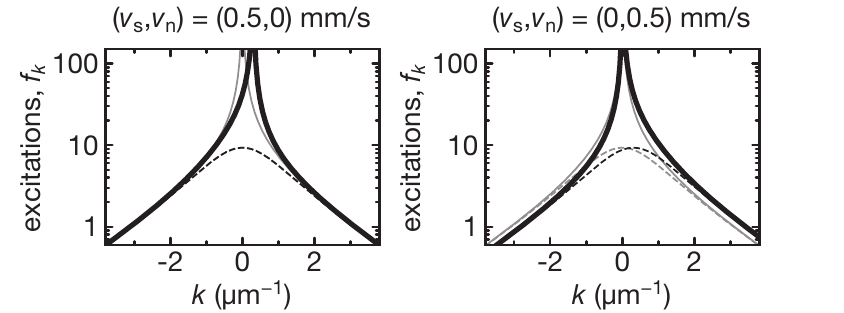}
\caption{
Momentum distribution of thermal excitations in moving two-fluid systems [Eq.~(\ref{eq_fk_moving_Landau})], for a weakly interacting Bose gas.
The black lines show the number of excitations per $k_z$-state based on the Bogoliubov dispersion for $^{39}$K at $k_x=k_y=0$, $T=100\,$nK, and $gn_{\rm BEC}/\kB=10\,$nK. 
The particle-like excitations have an equilibrium distribution in a frame moving with $\vn$ (dashed lines based on the Hartree--Fock approximation), while the phononic excitations mostly follow the motion of the BEC (not shown) with $\vs$.  The gray lines show the unshifted distributions.
}
\label{fig_SIexcitations}
\end{figure}

\subsection{Superfluid and normal densities}
\label{sect_defnormal}

The superfluid and normal component of a nonzero-temperature superfluid are typically associated with the condensed and thermal components, respectively, but in interacting systems this mapping does not fully hold. 

Landau~\cite{Landau_SI:1941} defined the normal density $\nn$ and the superfluid density $\ns \equiv n-\nn$ (where $n$ is the total density) through the Galilean invariance of an excitation spectrum, without reference to a condensate. Namely, for a normal velocity $\bm{\vn}$ (externally imposed \emph{e.g.}~by a moving wall with friction) in the frame where the superfluid velocity $\bm{\vs}=0$, the normal density $\nn$ is defined by equating the hydrodynamic normal mass current (particle mass $m$) to the total momentum of the excitations with microscopic momenta $\hbar \bm{k}$:
\vspace{-0.5em}
\begin{align}
m\nn {\bm \vn}  = \int \frac{{\rm d}\bm{k}^3}{(2\pi)^3} \hbar{\bm k}\,f({\bm k};{\bm\vn})\label{eq_def_nn} \,.
\end{align}

The distribution $f$ is
\vspace{-0.5em}
 \begin{align}
 f(\bm{k};\bm{\vn}) = \frac{1}{{\exp\left[\frac1{\kB T}\braces{\varepsilon(\bm{k})-\hbar \bm{k}\cdot\bm{\vn}}\right]-1}}\,,\label{eq_fk_moving_Landau}
 \end{align}
where $\bm{k},\bm{\vn}$, and the excitation energy $\varepsilon(\bm{k})$ are all measured in the frame of the superfluid. Note that $f$ is just the Bose distribution with excitation energies in the frame moving with $\bm{\vn}$\cite{Pethick_SI:2002} and that $|\bm{\vn}|$ needs to be smaller than the critical velocity, $v_{\textrm{c}}=\min(\varepsilon(k)/k)$, for $f$ to be finite. 

In general, combining Eqs.~(\ref{eq_def_nn}, \ref{eq_fk_moving_Landau}) shows that $\nn$ is a function of $\bm{\vn}$, but for $\vn\ll v_{\textrm c}$ the integral is linear in $\bm{\vn}$ so that $\nn$ is independent of it and given by: 
\vspace{-0.5em}
\begin{align}
\label{eq_popovnnns}
\nn &= -\frac{4\pi}{(2\pi)^3}\int {\rm d}k\, k^2 \frac{\hbar^2 k^2}{3m}\frac{\partial f_0(\varepsilon)}{\partial \varepsilon}\, ,
\end{align}
where $f_0(\varepsilon)$ is the usual Bose function ($\vn=\vs=0$).

In Fig.~\ref{fig_SIexcitations}, we illustrate the lab-frame distribution $\tilde{f}({\bm k};{\bm \vn},{\bm \vs}) = f(\bm{k}-m\bm{\vs}/\hbar;{\bm \vn}-{\bm \vs})$ for our typical experimental values. Effectively, those parts of the thermal distribution that react to a change in $\vs$ are part of the superfluid, while those that shift with $\vn$ form the normal component. For particle-like excitations, where $\varepsilon(\bm{k})= (\hbar k)^2/2m+gn$ [Hartree--Fock (HF) approximation], $ \tilde{f}(\bm{k};\bm{\vs},\bm{\vn})$ is merely shifted to be centered at $\bm{\vn}$, so all those are part of the normal fluid. Phononic excitations, however, are tied to the condensate, and thus the peak of the Bogoliubov-based distribution shifts with $\bm{\vs}$.

\subsection{Hydrodynamic equations}
\label{sec_hydro_SF}
Landau's two-fluid sound equations~\cite{Landau_SI:1941b} (see also \cite{Pitaevskii_SI:2016,Griffin_SI:2009})
\begin{align}
\partial_t^2 n &= \frac1{m} \nabla^2 p\,,\label{eqs_landau_sound}\\
\partial_t^2 \tilde{s} &= \frac{\nseq}{m\nneq}\tilde{s}_0^2\nabla^2 T\,,\nonumber
\end{align}
describe weak excitations in a nonzero-temperature superfluid. They relate fluctuations in density $n$ and entropy per particle $\tilde{s}$ to those of the pressure $p$ and temperature $T$ through the equilibrium superfluid and normal densities ($\nseq$,$\nneq$) and entropy per particle $\tilde{s}_0$. The properties of different superfluids enter from the equation of state via four thermodynamic derivatives:
\begin{subequations}
\label{eqs_Landau_linearhi}
\begin{align}
\delta p &= \drvTD{p}{n}{S,N}\delta n+
\drvTD{p}{\tilde{s}}{V,N}
\delta\tilde{s}\label{Ll8}\,,\\
\delta T &= \drvTD{T}{n}{S,N}\delta n +\drvTD{T}{\tilde{s}}{V,N}\delta\tilde{s}\label{Ll9}\,,
\end{align}
\end{subequations}
where $S$ and $N$ are, respectively, the entropy and atom number in a volume $V$.

In this section, we recast these equations in a way that 
emphasizes the crossover between the compressible and incompressible regime as a function of a single control parameter~$K$ [Eq.~(1) of the main paper].
Taking a step back, we start from Landau's general two-fluid equations for a dissipationless fluid in local thermal equilibrium with density $n=\ns+\nn$ and density current $\bm{j} = \ns \bm{\vs} + \nn \bm{\vn} = n\bm{v}$ at each point in space and time. For an entropy density $s$, pressure $p$, chemical potential $\mu$, and an external potential $V_{\rm ext}$: 
\begin{subequations}
\label{eqs_Landau}
\begin{align}
\partial_t n &=- \nabla\cdot\bm{j}\label{eq_W2}\,,\\
\partial_t \bm{v}_{\rm s}+\bm{v}_{\rm s}\cdot\nabla\bm{v}_{\rm s}
 &=-\frac1{m}\nabla\braces{\mu+ V_{\rm ext}}  \label{eq_W1}\,,\\
\partial_t  \bm{j} + \bm{w}
&=- \frac1{m}(\nabla p+n\nabla V_{\rm ext})\label{eq_W3}\,,\\
\partial_t s &=- \nabla\cdot(s\bm{\vn})\label{eq_W4} \,,
\end{align}
\end{subequations}
where 
\mbox{$\bm{w} = \sum_{\mu\nu}\hat{\bm{e}}_{\mu}\partial_{\nu} (\ns v_{{\rm s}\mu} v_{{\rm s}\nu} + \nn v_{{\rm n}\mu} v_{{\rm n}\nu})$}. To look at the linear response to a time-varying $V_{\rm ext}$, we split all (local) thermodynamic quantities into their equilibrium values and small fluctuations, assuming the form $A=A_0+\var[-0.1]{A}\,e^{i(\bm{k}\cdot\bm{r}-\omega t+\phi)}$; in cases where it is clear from the context that we refer to the equilibrium value, we omit the subscript $0$ to improve readability.

Linearizing the equations around static equilibrium ($\vneq=\vseq=0$) gives four dynamic equations:

\begin{subequations}
\label{eqs_Landau_linear}
\begin{align}
\omega {\var{\bm{v}_{\rm s}}}  &=\frac1{m}\braces{\bm{k}\var[-0.1]{\mu}+ \bm{k}\var{V_{\rm ext}}}\label{Ll1}\,,\\
\omega \var{n} &= \bm{k}\cdot\var{\bm{j}} \label{Ll2}\,,\\
\omega \var{\bm{j}} &= \frac1{m}\left(\bm{k}\var{p}+n_0\bm{k} \var{V_{\rm ext}}\right)\label{Ll3}\,,\\
\omega \var{s} &= s_0\bm{k}\cdot\var{\bm{v}_{\rm n}}\label{Ll4}\,,\\
\intertext{where}
\var{\bm{j}} &= \nseq\var{\bm{v}_{\rm s}}+\nneq\var{\bm{v}_{\rm n}}\label{Ll5}\,,\\
\var{s}&=n_0\var{\tilde{s}}+\tilde{s}_0\var{n}\label{Ll6}\,.
\end{align}
\end{subequations}
Using Eqs.~(\ref{eqs_Landau_linearhi},\,\ref{eqs_Landau_linear}) and the Gibbs--Duhem relation
\begin{align}
\var{\mu} &= -\tilde{s}_0\var{T}+\frac1{n_0}\var{p}\label{Ll7}\,,
\end{align}
one can solve for the oscillation amplitudes ($\delta \vs,\delta \vn, \delta j, \delta n, \delta s, \delta \tilde{s}, \delta T, \delta p, \delta \mu$).
Note that \mbox{$\delta \nn$ and $\delta \ns$} are not part of these equations and there are no separate continuity equations for $\nn$ and $\ns$. The superfluid and normal components locally convert into one another, and for dissipative systems this leads to two chemical potentials and an additional dynamic equation~\cite{Griffin_SI:2009}. Here we use the instantaneous-thermalization approximation, where $\delta \nn$ is set by $\delta n$, $\delta T$, and $\delta \mu$.

Restricting ourselves to 1D motion and using two-component vector notation for the \textit{relative} changes in a free system ($V_{\rm ext}=0$), we use Eqs.~(\ref{eqs_Landau_linear},\,\ref{Ll7}) to write:
\begingroup
\allowdisplaybreaks
\begin{align}
\pvec{\var{n}/n \\ \var{\tilde{s}}/\tilde{s}}
&\eqtext{\ref{eqs_Landau_linear}f} \overbrace{\pvec{1 &0\\-1 &1}}^{{\rm \bf M}_1}\pvec{\var{n}/n\\ \var{s}/s} 
\eqtext{\ref{eqs_Landau_linear}b,d} \frac{{k}}{\omega} {\rm \bf M}_1 \pvec{\var{{v}}\\ \var{{\vn}}}\nonumber\\
&\eqtext{\ref{eqs_Landau_linear}e}\frac{{k}}{\omega} {\rm \bf M}_1 \overbrace{\pvec{1&0\\ \frac{n}{\nn}&-\frac{\ns}{\nn}}}^{{\rm \bf M}_2}\pvec{\var{{v}}\\ \var{{\vs}}}\nonumber\\
&\eqtext{\ref{eqs_Landau_linear}a,c}\frac{k^2}{m\omega^2} {\rm \bf M}_1 {\rm \bf M}_2 \pvec{\var{p}/n\\ \var{\mu}}\nonumber\\
&\eqtext{\ref{Ll7}}\frac{k^2}{m\omega^2} {\rm \bf M}_1 {\rm \bf M}_2 {\pvec{1 &0\\1&-1}} \pvec{\var{p}/n\\ \var{T}\ \tilde{s}}\nonumber\\
&=\frac{k^2}{m\omega^2} \pvec{1 &0\\0&\frac{\ns}{\nn}} \pvec{\var{p}/n\label{eq_wave_2F}\\ \var{T}\ \tilde{s}}\,.
\end{align}
\endgroup
This is Eq.~(\ref{eqs_landau_sound}) in Fourier space.

We parametrize the system-specific derivatives [\mbox{Eqs.~(\ref{eqs_Landau_linearhi})}] using the classical speed of sound $\cc$ (density wave), a speed $b$ (entropy wave), and two dimensionless numbers $K$ and $J$:
\begingroup
\allowdisplaybreaks
\begin{align}
\label{eqs_defJKb}
\cc^2 &\equiv \frac1{m}\drvTD{p}{n}{S,N} = \frac1{mn\kappa_S}\,,\\
b^2 &\equiv \frac{\tilde{s}^2}{m}\drvTD{T}{\tilde{s}}{V,N} = \frac{Ts^2}{mnc_V}\nonumber\,,\\
K^2 &\equiv 1- \frac{c_V}{c_P} =\frac{\kappa_T}{c_P}T\drvTD{p}{T}{V,N}^2=\frac{\kappa_S}{c_V}T\drvTD{p}{T}{V,N}^2\nonumber\,,\\
J &\equiv - \frac{n}{\tilde{s}}\drvTD{\tilde{s}}{n}{T,N}
= \frac{1}{s}\drvTD{p}{T}{V,N}\nonumber
= 1-\frac1{\tilde{s}}\drvTD{S}{N}{T,V}\,.
\end{align}
\endgroup
Here we used a Maxwell relation and the standard expressions for the heat capacity difference and ratio to show different useful forms of these parameters. Note that $bJ = \cc K$, so only three of the four quantities are independent.
With these parameters Eqs.~(\ref{eqs_Landau_linearhi}) can be written as

\begin{align}
\pvec{\frac{\var{p}}{n}\\[8pt] \var{T}\ \tilde{s}} 
=m
\pvec{{\cc^2} & b \cc K\\[8pt] 
b\cc K & b^2} 
&\pvec{\frac{\var{n}}{n} \\[8pt] \frac{\var{\tilde{s}}}{\tilde{s}}},
\label{eq_Thermo_pT}
\end{align}
where we used $\drvTD{p}{\tilde{s}}{V,N}=-N\drvTD{T}{V}{S,N}=n^2\drvTD{T}{n}{S,N}$ and $\drvTD{T}{n}{\tilde{s}} = -\drvTD{\tilde{s}}{n}{T}\drvTD{T}{\tilde{s}}{n}=mJb^2/(n\tilde{s})$.

\vspace{0.5mm}
Inserting the r.h.s. of Eq.~(\ref{eq_Thermo_pT}) into Eq.~(\ref{eq_wave_2F}), we obtain 
\begin{align}
\frac{\omega^2}{k^2}\pvec{\frac{\delta n}{n} \\[8pt] \frac{\delta\tilde{s}}{\tilde{s}}}
&=
\pvec{
\cc^2  &  b^2J\\[8pt]
\frac{\ns}{\nn}b^2J  &  \frac{\ns}{\nn}b^2
}
\pvec{\frac{\delta n}{n} \\[8pt] \frac{\delta\tilde{s}}{\tilde{s}}}\label{eq_wave_c0bJ}.
\end{align}
This is a generalized eigenvalue equation for the (squared) speed of sound $c^2=\omega^2/k^2$, which can be written with two symmetric matrices [Eq.~(1) of the main paper]:
\begin{align}
c^2
\pvec{1 & 0\\
0 & \tfrac{\nn}{\ns}
}
\pvec{\frac{\delta n}{n} \\[8pt] \frac{\delta\tilde{s}}{\tilde{s}}} 
&=
b^2\pvec{
J^2/K^2                 &                  J  \\[8pt]
J  &  1
}
 \pvec{\frac{\delta n}{n} \\[8pt] \frac{\delta\tilde{s}}{\tilde{s}}}.\label{eq_1_in_SI}
\end{align} 
\indent $J$ is of order 1 for Helium-4, ultracold Bosons, and unitary Fermions at all $T< \Tc$; note that  $J=\kB\tilde{s}^{-1}\ll1$ for $T\gg\Tc$. 

Thus, the mode-structure of the two sounds is controlled by the heat-capacity ratio (via $K$) and the superfluid fraction (via $\nn/\ns$). All qualitative differences between materials are encoded in $K$, which varies between 0 and 1. The absolute sound speeds also depend on $b$, which varies widely for different materials and temperatures. 

\begin{figure}[!b]
\centering
\includegraphics[width=1\columnwidth]{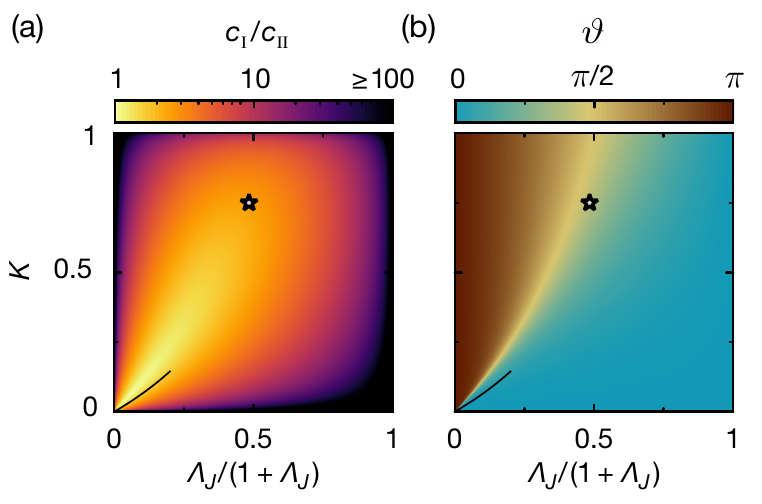}
\caption{
Structure of the solutions of the two-fluid hydrodynamic sound equations. (a) The ratio of sound speeds $c_{\rm I}/c_{\rm II}$, and  (b) the mixing angle $\vartheta$ in the ($\var{n}/n,\var{\tilde{s}}/\tilde{s}$) basis [see Eq.~(\ref{eq_general_modes_ns})] are uniquely determined by $\Lambda_J=J\sqrt{\nn/\ns}$ and $K$. Note that the x-axis runs from $T/\Tc=0$ to $1$ non-linearly.
The star marks the parameters for our measurements from Fig.~2(a) and Fig.~3 in the main paper, for which $\vartheta\sim\pi/2$ and the modes ($\var{n}/n,\var{\tilde{s}}/\tilde{s}$) are  strongly coupled. The black line corresponds to parameters valid for 3D superfluids with predominantly phononic excitations.
}
\label{fig_SIGensol}
\end{figure}

\subsection{General solution and limiting cases}
Equation (\ref{eq_1_in_SI}) is of the form  $\braces{{\rm \bf A}-{\rm \bf M}c^2}\bm{x}=\bm{0}$, with symmetric positive semi-definite matrices ${\rm \bf A}$ and ${\rm \bf M}$. The eigenvalues $c_i^2$ are thus real and non-negative. 
Written with a mixing angle $\vartheta\in[0,\pi]$ in analogy to the Bloch sphere: 

\begin{align}
\cot(\vartheta) &= \frac{J}{2K^2}\sqrt{\frac{\nn}{\ns}}-\frac1{2J}\sqrt{\frac{\ns}{\nn}}\,,\label{eq_def_theta_ns}
\end{align}
the speeds are 
\begin{align} 
c^2_{\rm I}&=b^2\braces{\frac{\ns}{\nn}+ J\sqrt{\frac{\ns}{\nn}}\cot(\vartheta/2)}\label{eq_generalFirstSound}\,,\\
c^2_{\rm II} &=b^2\braces{\frac{\ns}{\nn}- J\sqrt{\frac{\ns}{\nn}}\tan(\vartheta/2)}\label{eq_generalSecondSound}\,,
\end{align}
and the modes, each defined up to a prefactor, are 
\begin{align}
\ket{\rm I} &= \pvec{\cos(\vartheta/2)\\ \sqrt{\tfrac{\ns}{\nn}}\sin(\vartheta/2)}_{n,\tilde{s}},\, \nonumber
\\
\ket{\rm II} &= \pvec{-\sin(\vartheta/2)\\ \sqrt{\tfrac{\ns}{\nn}}\cos(\vartheta/2)}_{n,\tilde{s}}\label{eq_general_modes_ns},
\end{align}
where the subscripts highlight that these are written in the ($\var{n}/n,\var{\tilde{s}}/\tilde{s}$) basis.
Figure~\ref{fig_SIGensol} shows $\vartheta$ and $c_{\rm I}/c_{\rm II}$ for different $K$, $J$, and $\nn/\ns$.

It is instructive to consider 
different limiting cases:


\begin{itemize}
  \item For $K\ll1$, as for liquid {\bf Helium} at any $T<\Tc$, the density mode, $(1,0)_{n,\tilde{s}}$, and the entropy per particle mode, $(0,1)_{n,\tilde{s}}$, are decoupled. In this case 
  $c_{\rm I}=\cc$ and $c_{\rm II}=b\sqrt{\ns/\nn}$, with $b=\cc K/J\ll\cc$.
  For $T\to0$ ($\nn\to0$) most superfluids are incompressible \textbf{phonon gases}, with $J=1+d/2$ and $K=J\sqrt{\nn/(\ns d)}\to0$ in $d$ dimensions, so that $c_{\rm II}=\cc/\sqrt{d}$~\cite{Landau_SI:1941b}.

  \item For $\ns\ll\nn$, {\it i.e.}, $|T-\Tc|\ll\Tc$, we have a density mode with $c_{\rm I}=\cc$ for first sound and $c_{\rm II}=b\sqrt{(1-K^2)\ns/\nn}$ for second sound with eigenmode $(-K^2/J,1)_{n,\tilde{s}}$. This smoothly connects to the {\bf classical} situation above $\Tc$ (see Section~\ref{sec_si_damping}), where the second mode is a diffusive entropy mode with the same density admixture. 
  
  \item For $K,J\to1$, as in the \textbf{ideal gas}, the mode-couplings are strong. The first sound has $c_{\rm I}=\cc\sqrt{n/\nn}$ with the mode $(\nn/n,\ns/n)_{n,\tilde{s}}$. This is the mode of the normal velocity (derived in Section~\ref{subsec_nsbasis}), which is an entropy (per volume) wave. The second sound has $c_{\rm II}=\cc\tfrac{\ns}{n}\sqrt{(1-K^2)(1 - \tfrac{\nn}{n}(J^2-1) )}\to 0$ with the mode $(1,-1)_{n,\tilde{s}}$, corresponding to the motion of the superfluid.
\end{itemize}

\subsection{Normal-superfluid basis}
\label{subsec_nsbasis}

\begin{figure}[!t]
\centering
\includegraphics[width=1\columnwidth]{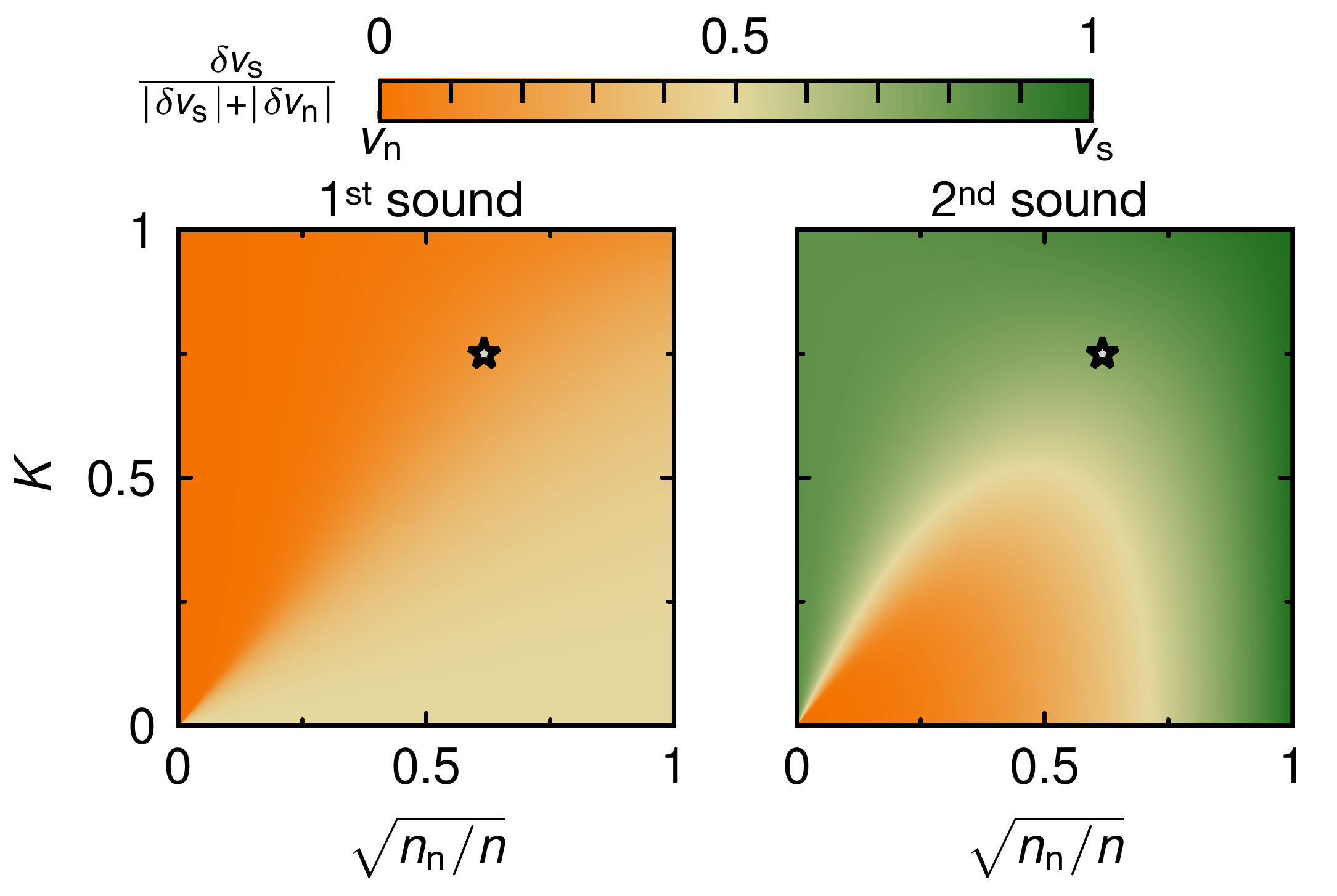}
\caption{Mode structure of the first and second sound in the normal-superfluid basis, shown as a function of $K$ and $\sqrt{\nn/n}$ for fixed \mbox{$J=1.2$}. In general, both modes correspond to motion of both fluids with an in-phase motion for the first sound and an out-of-phase motion for the second sound. Varying $T$ from $\Tc$ to $0$, the character of the second sound changes from a pure $\vs$ mode (green), to a mode with $\vn\gg\vs$ (orange). For the first sound, the normal velocity is always at least as large as the superfluid velocity. The star indicates parameters for our measurements from Fig.~2(a) and Fig.~3 in the main paper.
}
\label{fig_SIGensol_vs_vn}
\end{figure}

In our experiment, we measure the superfluid and normal velocities of the gas. 
Here we rewrite the two-fluid sound equation [Eq.~(\ref{eq_1_in_SI})] in the basis ($\nn\var\vn,\ns\var\vs$). As we show below, this is the eigenbasis of the compressible (ideal-gas) limit \mbox{($K\to1, J\to1$)}, where $|{\rm I}\rangle=(1,\ns/\nn)_{n,\tilde{s}}$ and $|{\rm II}\rangle=(1,-1)_{n,\tilde{s}}$.

We decompose an arbitrary perturbation ($\var{n}/n$, $\var{\tilde{s}}/\tilde{s}$) as
\begin{align}
\pvec{\frac{\delta n}{n} \\[8pt] \frac{\delta\tilde{s}}{\tilde{s}}} 
&= e_{\rm I}|{\rm I}\rangle
+e_{\rm II}|{\rm II}\rangle
=
\overbrace{ 
\pvec{1 & 1 \\[8pt] 
      \frac{\ns}{\nn} & -1}}^{\rm \bf R}
\pvec{e_{\rm I} \\[8pt] e_{\rm II}} \label{eq_introR},
\end{align}
so that using $\det({\rm \bf R})=-n/\nn$ we obtain 
\begin{align}
\pvec{e_{\rm I} \\[8pt] e_{\rm II}} &=
{\rm \bf R}^{-1} \pvec{\frac{\delta n}{n} \\[8pt] \frac{\delta\tilde{s}}{\tilde{s}}} 
= \frac{\nn}{n}
\pvec{1 & 1 \\[8pt] 
      \frac{\ns}{\nn} & -1}
\pvec{\frac{\delta n}{n} \\[8pt] \frac{\delta\tilde{s}}{\tilde{s}}} \nonumber\\
&\eqtext{\ref{eqs_Landau_linear}f}
\frac{\nn}{n}\pvec{\frac{\delta s}{s}\\[8pt]\frac{\delta n}{\nn} - \frac{\delta s}{s}}
\eqtext{\ref{eqs_Landau_linear}a,d}
\frac1{n c} \pvec{\nn\var{\vn} \\[8pt]  \var{j} - \nn\var{\vn}}\nonumber\\
&\eqtext{\ref{eqs_Landau_linear}e}
\frac1{n c}\pvec{\nn \var{v}_{\rm n} \\[8pt]  \ns \var{v}_{\rm s}}\label{eq_basis_idelgas}.
\end{align}

We see that the first sound of a highly compressible gas (where $e_{\rm I}\neq 0, e_{\rm II}=0$) is a wave where only the normal velocity is nonzero, while the second sound ($e_{\rm I}= 0, e_{\rm II}\neq0$) is a wave of the superfluid velocity. 

In general, there are no separate continuity equations for $\ns$ and $\nn$, so the theory makes no statement about their spatio-temporal changes. 
However, in the ideal-gas limit $\ns$ and $\nn$ do separately satisfy continuity equations, so that the first and second sound are also waves of the normal and superfluid \textit{density}, respectively\cite{Griffin_SI:2009}. This identification also extends close to the non-interacting limit, {\it i.e.}, to the weakly interacting Bose gas. 

Finally, we write the central two-fluid equation [Eq.~(\ref{eq_1_in_SI})] in the ($\nn\var\vn,\ns\var\vs$) basis by multiplying by ${\rm \bf R}^{\rm T}\nn c$ from the left, inserting ${\bf RR}^{-1}$, and using Eqs.~(\ref{eq_introR},~\ref{eq_basis_idelgas}) to obtain
 \begin{align}
b^2\Bigg[
\frac{n}{\nn}\pvec{1&0\\0&0} 
&+ \frac{\nn}{nK^2}\braces{J^2-K^2}\pvec{1&1\\1&1}\nonumber\\
&+ \frac{J-1}{n}\pvec{2\ns&\ns-\nn\\\ns-\nn&-2\nn}
\Bigg]
\pvec{\nn\delta \vn \\[8pt]  \ns\delta \vs} \nonumber\\
&=
c^2
\pvec{1 & 0 \\[8pt] 
      0 & \tfrac{\nn}{\ns}}
\pvec{\nn\delta \vn \\[8pt]  \ns\delta \vs}\label{eq_general_vn_vs_basis}.
\end{align}

We find the general solution of Eq.~(\ref{eq_general_vn_vs_basis}) in the  ($\nn\var\vn,\ns\var\vs$) basis and finally express it in the $(\var{\vn},\var{\vs})$ basis in terms of $\theta = \theta_K + \theta_J$:

\begin{align}
\cot(\theta_K) &= \frac12\frac{\Lambda_J^2+K^2/\Lambda_J^2+3K^2-1}{\Lambda_J(1-K^2)}\,,\nonumber\\
\tan\left(\frac{\theta_J}{2}\right) &= \frac{\Lambda_J(J-1)}{\Lambda_J^2+J}\,,
\end{align}
with eigenmodes
\begin{align}
\ket{\rm I} = \pvec{\cos\left(\theta/2\right)\\ \sqrt{\tfrac{\nn}{\ns}}\sin\left(\theta/2\right)}_{\vn,\vs}\nonumber,\nonumber\\
\ket{\rm II} = \pvec{-\sin\left(\theta/2\right)\\ \sqrt{\tfrac{\nn}{\ns}}\cos\left(\theta/2\right)}_{\vn,\vs},
\end{align}
where $\Lambda_J^2=J^2\nn/\ns$. In Fig.~\ref{fig_SIGensol_vs_vn}, we illustrate the mode structure in this basis for both sounds as a function of $K$ and $\sqrt{\nn/n}$, with $J=1.2$ (corresponding to our experiments).

\subsection{Weakly interacting Bose gas} 
For superfluid Bose gases in the compressible limit, as realized by our experiments, both $K$ and $J$ are close to $1$.
For sufficiently weak interactions the thermodynamic quantities can be calculated based on microscopic interactions. Provided that the interaction energy is small compared to the thermal energy, one can apply Hartree--Fock (HF) mean-field theory; see \hyperref[sec_bose_table]{Appendix~C} for a list of thermodynamic relations in this limit.

Inserting these HF-relations into Eq.~(\ref{eq_general_vn_vs_basis}) gives the sound equation up to linear order in $g$ (see also \cite{Griffin_SI:2009})
 \begin{align}
\pvec{\chi(z)\kB T+2g\nn & 2g\ns\\
2g\nn &g\ns}
\pvec{\delta \vn\\ \delta \vs} &= mc^2\pvec{\delta \vn\\ \delta \vs}\,.
\label{eq_couplingSF_oldEq2}
\end{align}
Here $\chi(z)=5 g_{5/2}(z)/(3 g_{3/2}(z))$ and the polylogarithms $g_{\alpha}(z)$ are evaluated at the fugacity $z=e^{\mu^*/(\kB T)}$, where $\mu^*=\mu-2gn$ is the interaction-corrected chemical potential. In the BEC phase, $\mu=gn_{\rm BEC}+2gn_T$, so $\mu^*=-gn_{\rm s}$, where we used that $n_{\rm BEC}=\ns$ in HF theory.
The normal and superfluid velocity modes are only weakly coupled provided $gn/(\kB T)\ll1$.

Up to first order in $g$, the first and second sound speeds can directly be read off from Eq.~(\ref{eq_couplingSF_oldEq2}):
\begin{align}
c_{\,\rm I}^{\rm (HF)}= \sqrt{\chi(z)\frac{\kB T}{m}+2\frac{g\nn}{m}}\,,\quad c_{\,\rm II}^{\rm (HF)} = \sqrt{\frac{g \ns}{m}}\,\label{eq_cH_SI}\,,
\end{align}
which is Eq.~(2) of the main paper. 
The eigenmode of the first sound is the normal velocity with a weak in-phase contribution of the superfluid velocity, $(\chi(z)\kB T, 2g\nn)_{\vn,\vs}$, and for second sound it is a superfluid motion plus a weak out-of-phase normal contribution,  $(-2g\ns, \chi(z)\kB T)_{\vn,\vs}$ [see Fig.~\ref{fig_SIGensol_vs_vn} at $K\approx1$].

For a description of weakly interacting Bose gases down to $T=0$, including the hybridization point\cite{Verney_SI:2015}, the phononic nature of the low-$k$ excitations needs to be included\cite{Pitaevskii_SI:2016}.

\subsection{Response function}  
\label{sec_response_function}
Experimentally, sound modes can be probed using weak external drives. Here we calculate the linear response to an oscillating gradient potential defined within a box of length~$L$
\begin{align}
V_\mathrm{ext}&=F_0\sin(\omega t)\,x\\
&=F_0\sin(\omega t)\sum_{q>0,\,\text{odd}}(-1)^{\tfrac{q-1}{2}}\frac{4}{L(qk)^2}\sin(qkx)\,\nonumber,
\label{eq_driveBox}
\end{align}
where $k=\pi/L$. We restrict the analysis to the lowest mode, $q=1$.

The external potential couples to the total density, which, within linear response, leads to a generalized (inhomogeneous) form of Eq.~(\ref{eq_wave_c0bJ}):
\begin{align}
\overbrace{m\pvec{
\cc^2-\frac{\omega^2}{k^2}  &  b^2J\\[8pt]
\frac{\ns}{\nn}b^2J  &  \frac{\ns}{\nn}b^2-\frac{\omega^2}{k^2}
}}^{B}
\pvec{\frac{\var{n}}{n} \\[8pt] \frac{\var{\tilde{s}}}{\tilde{s}}} 
=
\pvec{\delta V_{\omega,k} \\[8pt] 0},
\end{align}
where $\var{V}_{\omega,k}$ is the Fourier component of $V_\mathrm{ext}$.
The amplitudes of the density and entropy oscillations are then 
\begin{align}
\pvec{\frac{\var{n}}{n} \\[8pt] \frac{\var{\tilde{s}}}{\tilde{s}}} =
\mathbf{B}^{-1}\pvec{\delta V_{\omega,k} \\[8pt] 0}\,,
\end{align}
and the density-density response function, defined by $\delta n=\chi_{nn}\delta V_{\omega,k}$, is (see also \cite{Griffin_SI:2009}) 
\begin{align}
\chi_{nn}(w,k) &=n\left(\mathbf{B}^{-1}\right)_{11}
=n \frac{k^2}{m}\frac{\tfrac{\ns}{\nn}b^2k^2-\omega^2}{(\omega^2-c_\mathrm{I}^2k^2)(\omega^2-c_\mathrm{II}^2k^2)}\label{eq_responseFunction}.
\end{align}  
In general $\chi_{nn}$ has two resonances at $\omega_{\rm I,II}=c_{\rm I,II}k$ and a zero-crossing at $\omega=\sqrt{\ns/\nn}\,bk$. However, in the incompressible limit, where $c_{\rm II}=b\sqrt{\ns/\nn}$, the zero-crossing coincides with $\omega_{\rm II}$; this is the result familiar from Helium, that the second sound cannot be excited and probed via the density.

Including weak damping modifies $\chi_{nn}$ to~\cite{Hohenberg_SI:1965}
\begin{align}
\chi_{nn} &= n\frac{k^2}{m} \frac{\tfrac{\ns}{\nn}b^2k^2-\omega^2-ik^2\omega h_1}{\braces{\omega^2-c_{\rm I}^2k^2+i D_{\rm I}k^2\omega}\braces{\omega^2-c_{\rm II}^2k^2+iD_{\rm II}k^2\omega}}\label{eq_response_nn_withDamping}\nonumber,\\
\end{align}
where $D_{\rm I}$ and $D_{\rm II}$ are the (weak) diffusivities, and $h_1$ is of the same order as the diffusivities. 

One can rewrite Eq.~(\ref{eq_response_nn_withDamping}) as a sum of two resonances:
\begin{align}
\chi_{nn} & = \frac{nk^2}{m(c_{\rm I}^2-c_{\rm II}^2)}\times\label{eq_response_nn_withDamping2}\\
&\left(\frac{(\tfrac{\ns}{\nn}b^2-c_{\rm I}^2)-i\omega D_{\rm I}h_2}{\omega^2-c_{\rm I}^2k^2+iD_{\rm I}k^2\omega}\right.
-\left.\frac{(\tfrac{\ns}{\nn}b^2-c_{\rm II}^2)-i\omega D_{\rm II}h_2}{\omega^2-c_{\rm II}^2k^2+iD_{\rm II}k^2\omega}\right), \nonumber
\end{align}
with $h_2=1+([D_{\rm I}-D_{\rm II}]\frac{\ns}{\nn}b^2+[c_{\rm I}^2-c_{\rm II}^2]h_1)/(c_{\rm I}^2 D_{\rm II}-c_{\rm II}^2 D_{\rm I})$.
Note that $i\omega D_{\rm I,II} h_2$ are typically negligible around the resonances.

Finally, the experimentally relevant conductivity is given by $\sigma=i \omega n \chi_{nn}$.

\begin{figure}[!t]
\centering
\includegraphics[width=0.99\columnwidth]{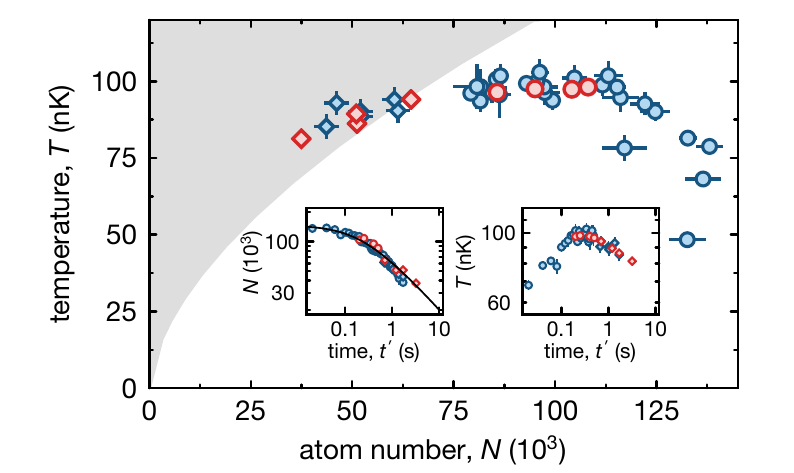}
\caption{
Thermodynamic evolution of our gas.
We plot $T(N)$ measured after increasing $a$ to \mbox{$480\,a_0$}; the insets show the corresponding time evolution.
The red symbols mark measurement conditions from Fig.~2 of the main text, while the blue symbols show reference measurements where we do not excite sound modes.
For thermal clouds (diamonds), we obtain $T$ by fitting the momentum distribution~\cite{Gaunt_SI:2013}.
For partially condensed clouds (circles), we measure the condensed fraction and $T$ using a BEC filtering technique based on Bragg pulses \cite{Gerbier_SI:2004c,Lopes_SI:2017b}.
The indicated transition to the thermal phase (gray shading) assumes an ideal-gas $\Tc$ (see Section~\ref{sect_critical_T}). 
The loss dynamics $N(t')$ are captured by a three-body rate $\dot{N}\propto N^3$ (black line), while $T(t')$ shows rapid heating to a trap-depth-limited $T\approx 100$\,nK, where evaporative cooling balances the heating caused by the loss processes. Since the three-body rate is $\propto N^3$ and the evaporative cooling rate is $\propto N^2$, the temperature slowly drops at long times.
}
\label{fig_SIref}
\end{figure}

\vspace{-1em}
\section{Measurement and analysis techniques}
\label{sec_exp_protocol}

\subsection{Realizing a hydrodynamic two-fluid system}

To produce hydrodynamic Bose gases at different $T/\Tc$, we begin with weakly interacting quasi-pure Bose--Einstein condensates of $N\approx1.5 \times 10^5$ atoms at $a\approx 200\,a_0$~\cite{Eigen_SI:2016}. We raise $a$ to $480(20)\,a_0$ in 50\,ms, which also enhances three-body recombination and the associated heating, as summarized in Fig.~\ref{fig_SIref}. 
The main panel shows the behavior of $T(N)$ as we vary the time $t'=t_{\rm h}+t$, where $t_{\rm h}$ is the hold time before we excite the cloud, and $t$ is the excitation time. To vary $T/\Tc$ for different measurements (see Fig.~2 in the main paper), we vary $t_{\rm h}$. We observe no significant difference in $N(t')$ and $T(t')$ with or without shaking (see Fig.~\ref{fig_SIref}).

Note that the elastic collision rate required to establish local thermodynamic equilibrium scales as $\Gamma_2\propto n a^2$, while inelastic losses close to a Feshbach resonance scale as $\Gamma_3\propto (n a^2)^2$ (up to corrections from Efimov physics\cite{Zaccanti_SI:2009}). At fixed $\Gamma_2$ (required to achieve a short $\mfp$), one can therefore not readily reduce $\Gamma_3$. Moreover, since $\mfp \ll L$, the loss products (with energies of $\sim 20\,\umu$K) do not immediately leave the system but undergo re-scattering, which heats the system. We reduce this effect by using an elongated box geometry.

\begin{figure}[!t]
\centering
\includegraphics[width=0.99\columnwidth]{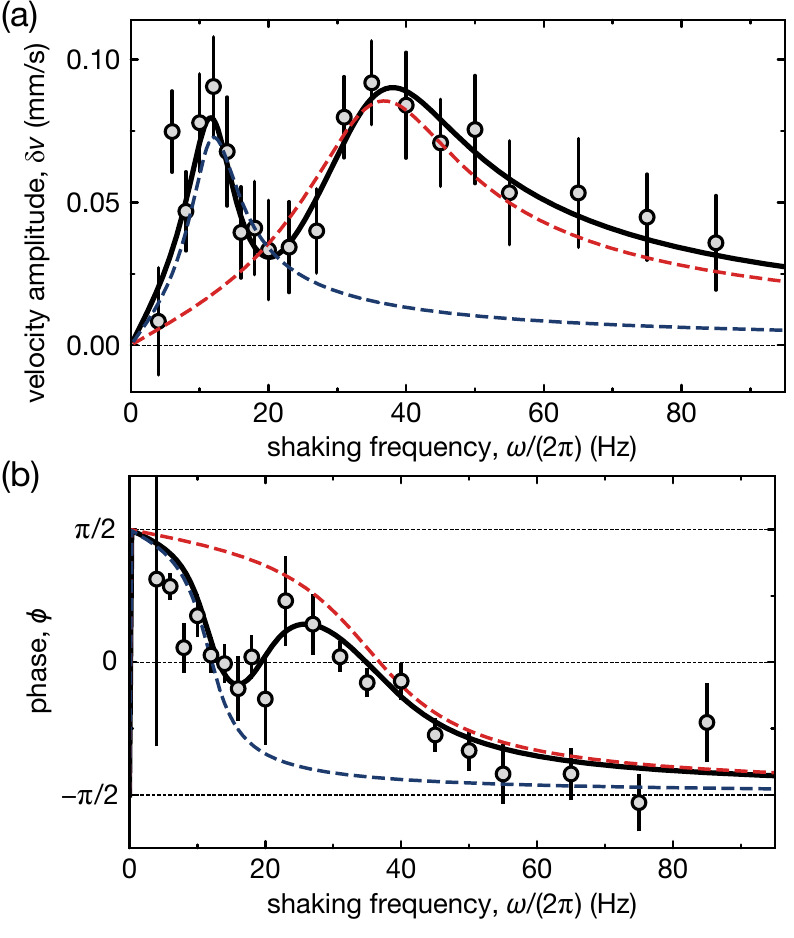}
\caption{
Measured (complex) velocity response, shown here for $T/\Tc=0.81$, $N=95 \times 10^3$ from Fig.~2(b) of the main text. From the instantaneous center-of-mass velocity $v(t)=\var v \sin (\omega t+\phi)$, we extract (a) the amplitude $\var v$ and (b) the phase $\phi$ (relative to the drive).
We fit the amplitudes using the (complex) sum of two velocity response functions [black lines, see Eq.~(\ref{eq_fitfunction_2L})] and see that the phase of the response is captured by the same parameters. The dashed red (blue) lines show the individual responses for the first (second) sound.}
\label{fig_SIfit}
\end{figure}

\subsection{Quadrants of the velocity response}
\label{sec_velquad}
To extend our measurements of the center-of-mass velocity response to many shaking frequencies $\omega$ and reduced temperatures $T/\Tc$, without mapping out the full time evolutions [as done in Fig.~2(a) of the main paper], we simply measure the four quadrants of the quasi-steady-state oscillation. Specifically, we measure at four quarter-period intervals around a mean shaking time $t$ of 200\,ms, {\it i.e.}, measurement $i\in\{1,2,3,4\}$ is performed by shaking for a time $t_i=200\,\textrm{ms} + 0.25\,(i-2.5)\, 2\pi/\omega$. Assuming a steady-state velocity $v_i=v_0 + \var v \sin(\omega t_i+\phi)$:

\begin{align}
 \delta v^2 &= \left[(v_1-v_3)^2+(v_2-v_4)^2\right]/4\,,\label{eq_quadrants}\\
 \tan\phi &= (v_1-v_3)/(v_2-v_4)\,,\nonumber
\end{align} 
gives $|\sigma| \propto \var v/F_0$ and the phase $\phi$ of the complex $\sigma(\omega)$, while the average velocity $v_0 = {(v_1+v_2+v_3+v_4)}/4$ should be zero; in practice a nonzero measured $v_0$ can arise due to residual magnetic field gradients during time-of-flight, but this does not affect the results in Eq.~(\ref{eq_quadrants}).
We repeat each $v_i$ measurement 6 - 12 times. In Fig.~\ref{fig_SIfit} we show a typical example of the extracted $\var v(\omega)$ and $\phi(\omega)$.

\subsection{Two-fluid response fit functions}

Within the two-fluid model, the optical conductivity $\sigma= nv/F$ can be approximated by a sum of two resonances (Section~\ref{sec_response_function}):
\begin{align}
\sigma=i\omega\braces{\frac{ A_{\rm I}}{\omega_{\rm I}^2-\omega^2+2i\gamma_{\rm I}\omega}
+ \frac{A_{\rm II}}{\omega_{\rm II}^2-\omega^2+2i\gamma_{\rm II}\omega}}.
\label{eq_fitfunction_2L}
\end{align}
As shown in Fig.~\ref{fig_SIfit}(a), we fit $\var v$ using the functional form of $|\sigma|$ to extract the resonance frequencies $\omega_{\rm I,II}$ and the damping coefficients $\gamma_{\rm I,II}$.

The same fit parameters also capture the phase response [see Fig.~\ref{fig_SIfit}\,(b)].
To compare measurements at different $T/\Tc$ and $N$, we show $\sigma\propto N v/F$ in Fig.~2 of the main paper.

\subsection{Calculating $\ns$ and $\nn$ for our gas}

\label{sect_Thermodynamics_Box}

\begin{figure} 
\centering
\includegraphics[width=0.99\columnwidth]{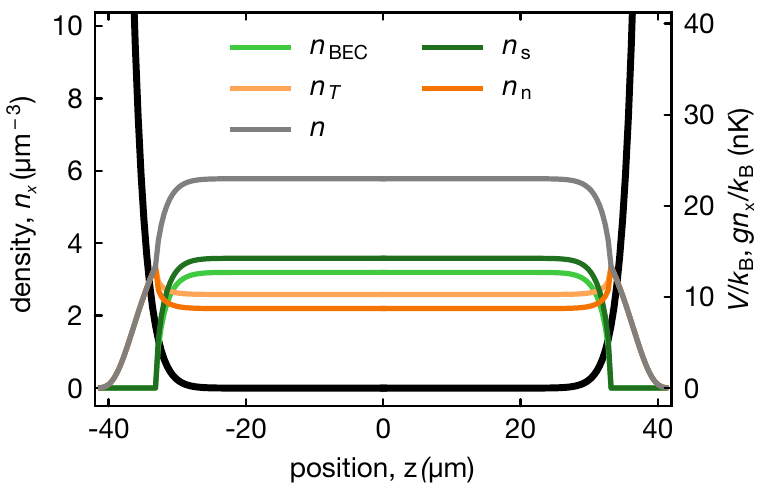} 
\caption{Calculated axial density distributions for \mbox{$N=105\times 10^{3}$} atoms at $T=97\,$nK and $a=480\,a_0$, obtained from the Popov calculation in LDA assuming a power-law potential (black line, for details see text). The total density $n=n_{\rm BEC}+n_T=\ns+\nn$ splits into the BEC and thermal densities, $n_{\rm BEC}$ and $n_{T}$, or the superfluid and normal densities, $\ns$ and $\nn$. For the hydrodynamic processes, $\ns$ and $\nn$ are the relevant variables.
For our calculations of the sound speeds and modes, we use the central density from this calculation but otherwise assume an ideal box.}
\label{fig_SIbox}
\end{figure}

Our optical trap is made from a pseudo Laguerre--Gaussian beam (with $8\times2\pi$ phase winding around the optical axis) and two light sheets~\cite{Gaunt_SI:2013}.
For most purposes it can be treated as a perfect cylindrical box with radius \mbox{$R=9.2(5)\,\umu$m} and length \mbox{$L=70(2)\,\umu$m}. 
However, to accurately capture thermodynamic behavior, we model the potential far below the trap depth by a power law in cylindrical coordinates (see Fig.~\ref{fig_SIbox}):
\begin{align}
V_{\rm box}(r,z)=\braces{\frac{r}{R}}^{\beta_r}V_r+\braces{\frac{2z}{L}}^{\beta_z}V_z.\label{eq_box_potential}
\end{align}
We calibrate the effective box size ($R$, $L$) and the steepness of the walls ($\beta_r$, $\beta_z$) based on a set of measurements: in-situ density profiles, scaling of $\Tc$ with $N$, and scaling of the Bogoliubov eigenmode frequency of a quasi-pure BEC for varying trapping laser powers (for methods see\cite{Schmidutz_SI:2014,Navon_SI:2018}). The measurements give combined best estimates $L=70(2)\,\umu$m, $R=9.2(5)\,\umu$m, $\beta_r=13(3)$, and $\beta_z=20(6)$, where $V_r=V_z=\,\kB \times20\,$nK. The corresponding thermodynamic scaling parameter\cite{Pethick_SI:2002} is $\alpha=\frac32+\frac1{\beta_z}+\frac2{\beta_r}=1.70(3)$. 
For the additional box used in the crossover measurements (see Fig.~4 of the main text), we find $L=50(2)\,\umu$m, \mbox{$R=15(1)\,\umu$m} and a similar $\alpha$.

For calculating  $\ns$, $\nn$, and $n$ at a given $N$, $T$, and $g=4\pi \hbar^2a/m$, an ideal-gas treatment is not sufficient, due to the relatively strong interactions [$\mu/(\kB T)$ up to $0.35$, $n^{1/3}a$ up to $0.05$]. Instead, we numerically solve the self-consistent Popov equations (see also \hyperref[sec_lk_table]{Appendix~B}), including the effects of a power-law potential via the local density approximation (LDA);
the results of such a calculation for our experimental parameters are illustrated in Fig.~\ref{fig_SIbox}. To calculate the expected sound speeds [Eq.~(2) of the main paper], we use the calculated $\ns$ and $\nn$ at the center of the box.
We also note that in the radial direction the admixture of inhomogeneous sound modes is suppressed, because the small box diameter leads to large radial-excitation energies. 

\subsection{Critical temperature $\Tc$}
\label{sect_critical_T}
An ideal gas condenses at a critical temperature \mbox{$T_{{\rm c},0}\sim N^{1/\alpha}$}. For an interacting gas in an inhomogeneous potential, one expects both a negative mean-field (MF) shift of $\Tc$ and a positive beyond-MF one~\cite{Smith_SI:2011b}.
For our system parameters and our nearly perfect box, we estimate a MF shift of $-4\%$ (LDA-Popov calculation), while the beyond-MF one is expected to be about $+ 3\%$\cite{Pilati_SI:2008}. We thus expect the two to almost cancel, and since our measurements do not resolve $\Tc$-shifts of order $1\%$, we simply assume $\Tc=T_{{\rm c},0}$.

\subsection{Measuring the momentum distribution} 
To measure the axial 1D momentum distribution $n_k(t)$ (shown in Fig.~3 in the main paper) we take absorption images of our samples perpendicular to the box axis, following a time-of-flight ballistic expansion of duration $t_{\rm ToF}$.
For this, we quench $a\rightarrow 0$ in $\sim2$\,ms, concurrently with the release of the atoms from the trap. The images yield the line-of-sight integrated momentum distribution, which we further integrate perpendicular to the driving direction $\hat{z}$. To improve our momentum resolution and range, we extract $n_k$ for $k>2\,\kunit$ using $t_{\rm ToF}=30$\,ms, and for $k<3\,\kunit$ using $t_{\rm ToF}=45$\,ms, and then smoothly combine the two measurements in the overlap region.

\subsection{Extracting superfluid and normal-fluid velocities} 
Here we explain our fit-free extraction of $\vs$ and $\vn$ from the measured $n_k(t)$ shown in Fig.~3 of the main paper.

\begin{figure}[t]
\centering
\includegraphics[width=0.99\columnwidth]{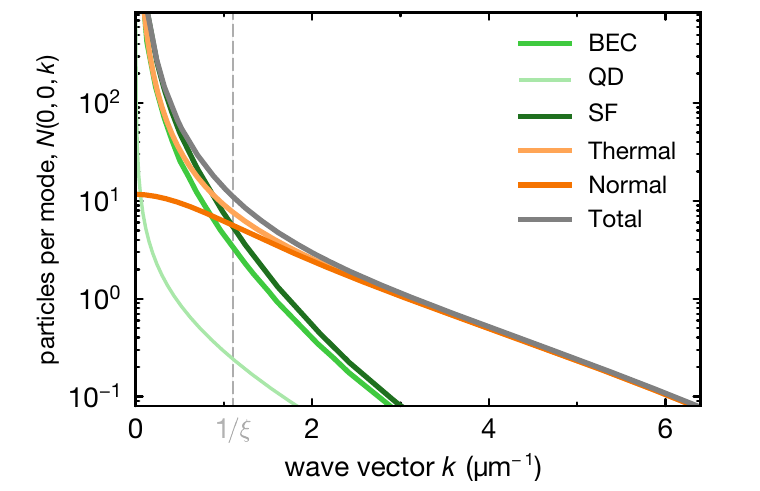}
\caption{Calculated momentum-space distribution in the axial direction [$\bm{k}=(0,0,k)$] for $N=105\times 10^{3}$ atoms in our box at $T=97\,$nK and $a=480\,a_0$. The condensate distribution, calculated from the Gross--Pitaevskii equation, has fast oscillations in $k$-space with period $2\pi/L=0.09\,\umu$m$^{-1}$ due to the finite box size; for clarity, we show an average over these oscillations (BEC, green). The excitations are calculated within the Popov approximation based on the density at the trap center. 
The thermal depletion (light orange) consists of particle-like excitations, which form the normal component (orange), and phononic excitations, which together with the BEC form the superfluid (dark green). The thermal distribution has Gaussian tails at momenta larger than those shown. The interaction-driven quantum depletion (QD, light green) is negligible for our parameters.}
\label{fig_SI_SF_N_k-space} 
\end{figure}

The low-$k$ dynamics involve a complex interplay of normal and superfluid components: even an equilibrium BEC has a nonzero $k$-space width and moreover a significant number of thermal excitations are phonons, which have $k\lesssim 1 / \xi = \sqrt{2 m g \nBEC} / \hbar$ and move partly with the superfluid and partly with the normal gas~\cite{Pitaevskii_SI:2016}.
However, at $k \gg 1/\xi$ there are only normal atoms  (see Fig.~\ref{fig_SI_SF_N_k-space}), and we directly extract $\vn$ from this part of the $n_k(t)$ distribution.

In analogy with a single-component classical gas in local equilibrium, we assume that the momentum distribution of the normal component is given by the equilibrium distribution $n_{{\rm n},k,0}(k)$ in the frame moving with $v\n(t) = \hbar k\n(t)/m$:
\begin{align}
 n_{k}(t) = n_{{\rm n},k,0}\left(k-k\n(t)\right)\, \text{for }|k|\geq \kc \,,
 \label{eq_nkt_kc}
\end{align}
where we choose $\kc=1.7\,\umu$m$^{-1}$, which is significantly larger than $1/\xi=1.1\,\umu$m$^{-1}$ but small enough to retain a good signal-to-noise ratio (see Fig.~\ref{fig_SI_SF_N_k-space} and also Fig.~3). 

We start by calculating the first moment of $n_k(t)$ only for $|k|>\kc$:
\begin{align}
\avg{k}_>
&\equiv N_>^{-1}\ \ {\int\limits_{\mathclap{|k|>k_{\rm c}}} \,k n_k(t)\,{\rm d}k}\,,\label{eq_def_slit} 
\end{align}
where the atom number $N_>\equiv\int\limits_{\mathclap{|k|>\kc}} \,n_k(t)\,{\rm d}k$ is almost independent of time for small $k_{\rm n}$, {\it i.e.}, $N_>\approx\int\limits_{\mathclap{|k|>\kc}} \,n_{{\rm n},k,0}(k)\,{\rm d}k$.
The moment $\avg{k}_>$ can be directly extracted from the experimental data, but is not equal to $k\n$, as we show below.

To get to $k\n$, we next introduce $k'=k-k\n(t)$ in Eq.~(\ref{eq_def_slit}), which leads to an asymmetric integration range that we split into $|k'|>k_{\rm c}$ and two boundaries

\begin{align}
\int\limits_{\mathclap{|k|>k_{\rm c}}} \,k n_k(t)\,{\rm d}k&=
\hspace{3mm}\int\limits_{\mathclap{|k'|>k_{\rm c}}}{[k'+k_{\rm n}(t)]} n_{{\rm n},k,0}(k')\,{\rm d}k'\\
&\phantom{=}-\int\limits_{-k_{\rm c}-k_{\rm n}(t)}^{-k_{\rm c}}{[k'+k_{\rm n}(t)]} n_{{\rm n},k,0}(k')\,{\rm d}k'\nonumber\\
&\phantom{=}+\int\limits_{k_{\rm c}-k_{\rm n}(t)}^{k_{\rm c}}{[k'+k_{\rm n}(t)]} n_{{\rm n},k,0}(k')\,{\rm d}k'.\nonumber
\end{align}

For $k\n$ much smaller than the width of the distribution, we expand
$n_{{\rm n},k,0}$ around $\pm k_{\rm c}$, and use the symmetry of $n_{{\rm n},k,0}$ to find
\begin{align}
\avg{k}_>
&=k_{\rm n}(t)\left(1 + 2 \frac{k_{\rm c} n_{{\rm n},k,0}(k_{\rm c})}{N_{{\rm n},|k|>k_{\rm c}}}\right)+\mathcal{O}\braces{k_{\rm n}^3}\label{eq_partial_COM_kLarge}.
\end{align}

Inverting Eq.~(\ref{eq_partial_COM_kLarge}) gives
\begin{align}
\vn(t) &= \frac{\hbar}{m}\avg{k}_{>}\braces{1+2\frac{k_{\rm c} n_{{\rm n},k,0}(k_{\rm c})}{N_{{\rm n},|k|>k_{\rm c}}}}^{-1} \, ,
\end{align}

where now all the quantities on the r.h.s.~can be directly extracted from the measured $n_k(t)$.

To extract $\vs(t)$, we additionally use the first moment of the total $n_k$ distribution, $\avg{k}$, and note that the corresponding velocity, $v(t) = \hbar \avg{k}/m$, is given by the sum of the superfluid and normal currents:
\begin{align}
v(t) = \frac{\hbar}{mN} \int k n_k(t)\,{\rm d} k=  \frac{N_{\rm s}\vs(t)+N_{\rm n}\vn(t)}{N_{\rm s}+N_{\rm n}}\, ,\label{eq_v_total}
\end{align} 
where $N\s$ and $N\n$ are, respectively, the total numbers of atoms in the superfluid and normal components. We thus have
\vspace{-0.5em}
\begin{align}
\vs(t) &= \frac{1}{N\s}\braces{Nv(t)- N\n\vn(t)} \, ,\nonumber
\end{align}
where $N\s$ and $N\n$ cannot be directly extracted from $n_k(t)$; instead we calculate them from the thermodynamics measurements discussed in Section~\ref{sect_Thermodynamics_Box}.

\subsection{Landau--Khalatnikov damping}
\label{sec_belowTc}

For a superfluid with hydrodynamic damping, the Landau--Khalatnikov equations~\cite{Khalatnikov_SI:1965} (see also \hyperref[sec_lk_table]{Appendix~A}) expand the dissipationless Landau equations by including a total of six transport coefficients (heat conductivity $\kappa$, shear viscosity $\eta$, and four secondary viscosities $\zeta_{1,2,3,4}$) that dictate the dissipative processes.
The expressions for $\gamma_{\rm I,II} = D_{\rm I,II}k^2/2$ were derived by Hohenberg and Martin~\cite{Hohenberg_SI:1965}.
Written with the sound speeds $c_{\rm I,II}$ defined by Eqs.~(\ref{eq_generalFirstSound},\,\ref{eq_generalSecondSound}) and notation from Eqs.~(\ref{eqs_defJKb}), one solves

\begin{align}
D_{\rm I}+D_{\rm II} &=  \frac{\kappa}{c_V}
+\frac{\frac43\eta+\zeta_2+\zeta_3m^2\ns n -(\zeta_1+\zeta_4)m\ns}{m\nn},\nonumber\\
c_{\rm I}^2D_{\rm II}+c_{\rm II}^2D_{\rm I} &= \cc^2\frac{\kappa}{c_P}+ \frac{\ns}{n}\frac{\frac43\eta+\zeta_2}{m\nn}\braces{b^2+\cc^2-2Jb^2}\nonumber\\
&\phantom{=}\hspace{2mm} +\zeta_3\frac{\ns}{\nn}\cc^2-(\zeta_1+\zeta_4)\frac{\ns}{\nn}\braces{Jb^2+\cc^2}\, \label{eq_chinn_withDamping}
\end{align}
for $D_{\rm I}$, $D_{\rm II}$.

Since the secondary viscosities scale as $\sim g^2$ in a weakly interacting gas, we set \mbox{$\zeta_1=\zeta_2=\zeta_3=\zeta_4 = 0$}.
Inserting the thermodynamic relations of the weakly interacting Bose gas (\hyperref[sec_bose_table]{Appendix~C}) and the mean-field results of the transport coefficients from\cite{Nikuni_SI:2001}, we can estimate the expected damping for our measurements below $\Tc$. %
\\
\\
\indent As an example, we provide the calculated quantities for \mbox{$T = 97(3)$ nK}, \mbox{$a = 480(20)\,a_0$}, and \mbox{$N = 105(3) \times 10^3$}:

At the trap center, we find \mbox{$n = 5.8(2) \times 10^{18}$ m$^{-3}$}, a condensed fraction $n_{\rm BEC}/n=55(3)\%$ and a superfluid fraction $\ns/n=62(3)\%$. As shown in Section~\ref{sect_Thermodynamics_Box}, these central values accurately describe most of the trap volume. Due to some additional thermal particles at the edge of the trap, $N_{\rm s}/N=43(3)\%$ is lower.

For the central values, we calculate (within Hartree--Fock theory) the heat capacities $c_V/n = 0.93\,\kB$ and $c_P/n = 2.1\,\kB$, the parameters  $K=0.75$, $J=1.20$, and $b=3.2\,$mm\,s$^{\text{--}1}$, the predicted sound speeds $c_{\rm I} = 5.0$ mm\,s$^{\text{--}1}$, $c_{\rm II} = 1.7$ mm\,s$^{\text{--}1}$, the heat conductivity $\kappa/\kB = 170 \times 10^{9}$\,(m\,s)$^{\text{--}1}$, and the shear viscosity $\eta/m = 42 \times 10^{9}$\,(m\,s)$^{\text{--}1}$.

This yields the hydrodynamic damping $\gamma_{\rm I} = k_L^2 D_{\rm I} / 2 =2 \pi \times 6.5$ Hz and \mbox{$\gamma_{\rm II} = k_L^2 D_{\rm II}/2 = 2 \pi \times 2.8$ Hz}.
\\
\\
\indent In the experimentally relevant range $N=\{80-105\}\times 10^3$, the Landau--Khalantinkov $\gamma_{\rm II}$ varies only slightly and we find an average $\gamma_{\rm II}\approx 2\pi \times 2.2$~Hz from this theory, consistent with our experimental results.

\section{Dissipative hydrodynamics above $\Tc$}
\label{sec_si_damping}
Here we extend our theoretical analysis to consider damping of sound in thermal gases, and derive the predictions shown in Fig.~4 of the main paper; in particular, that the speed of sound decreases with increasing damping for a wave with fixed wavelength.

The theory presented in Section~\ref{sec_si_theory} assumes instantaneously established local thermodynamic equilibrium at any ($\bm{r},t$), {\it i.e.}~that the particles are distributed according to the equilibrium distribution $f_0$ in the frame moving with $v(\bm{r},t)$.
In reality local equilibration relies on collisions and is not instantaneous. The deviations from the local equilibrium are at the heart of damping and, to lowest order in the local $\var{f}=f-f_0$, these effects are included in the hydrodynamic equations via the shear and bulk viscosity ($\eta$, $\zeta_2$) and heat conductivity ($\kappa$).  
All these transport processes are microscopically controlled by the collision rate. For very low collision rates, $\var{f}$ becomes large and the hydrodynamic description breaks down. We solve the hydrodynamic sound equations above $\Tc$ for arbitrary strength of the transport coefficients, assuming they hold even when approaching the collisionless regime.

\subsection{Damped sound waves}
We start with the Navier--Stokes equation~\cite{Landau_SI:1987} and loosely follow~\cite{Temkin_SI:1981}. We linearize the equations around a static equilibrium using the density ${n}$, velocity $v$, temperature ${T}$, entropy per particle ${\tilde{s}}$, and pressure ${p}$ as five space- and time-dependent variables (for notation see Section~\ref{sec_hydro_SF}). The three linearized dynamical equations are:

\begin{subequations}
\label{eqs_Classicalwaves}
\begin{align}
\partial_t  n &=- n_0\nabla \bm{v}\label{eq_Q1}\,,\\
mn_0\partial_t \bm{v} &= -\nabla  p
          +\eta\nabla^2 \bm{v}+(\eta/3+\zeta_2)\nabla\braces{\nabla\cdot \bm{v}}\label{eq_Q2},\\
n_0\partial_t \tilde{s} &= \frac{\kappa}{T_0}\ \nabla^2 T\label{eq_Q3}\,,
\end{align}
\end{subequations}
while the thermodynamic relations between $p,T,n,\tilde{s}$ are given by Eq.~(\ref{eq_Thermo_pT}).

In a single-component Bose gas above $\Tc$, the bulk viscosity vanishes ($\zeta_2=0$). 

We restrict ourselves to plane waves and 1D variations, {\it e.g.}
 \begin{align}
 n=n_0+\var{n}e^{i(kx-\omega t)}\, ,
\end{align}
and from Eqs.~(\ref{eqs_Classicalwaves}) obtain
\begin{subequations}
\label{eqs_ClassicalwavesB}
\begin{align}
\omega \delta n &= n_0 k \var{v}\label{eq_Q1b}\,,\\
\braces{mn_0\omega+i\frac43\eta k^2} \var{v} &= k \var{p}
           \label{eq_Q2b}\,,\\
n_0\omega \var{\tilde{s}} &= -i\frac{\kappa}{T_0}\ k^2\var{T}\,.\label{eq_Q3b}
\end{align}
\end{subequations}
The main effect of the nonzero transport coefficients is that density and pressure, as well as entropy and temperature, do not oscillate in phase. A change of density is also a change of the volume per particle, so focusing on a single point in space, a sound wave maps out an ellipse in $(p,V)$-space (as compared to a line in the dissipationless case). This means that the wave does work and thus loses energy.

We solve Eqs.~(\ref{eqs_ClassicalwavesB}) for the speed and damping of sound by eliminating the velocity $\var{v}$ and expressing $\var{p}$, $\var{T}$ in terms of $\var{n}$, $\var{\tilde{s}}$ (in analogy with the calculation in Section~\ref{sec_si_theory}B).
This leads to an analog of Eq.~(1) in the main text, but above $\Tc$ and including damping:

\begin{align}
\frac{\omega^2}{k^2}\pvec{\frac{\delta n}{n} \\[8pt] \frac{\delta\tilde{s}}{\tilde{s}}}
&=
\pvec{
\cc^2- i \frac43\frac{\eta\omega}{mn}   &  b^2J\\[8pt]
-i\frac{\kappa\omega}{c_V}J  &  -i\frac{\kappa\omega}{c_V}
}
\pvec{\frac{\delta n}{n} \\[8pt] \frac{\delta\tilde{s}}{\tilde{s}}}\label{eq_matrix_damping}.
\end{align}
This system of equations has nonzero solutions for $\var{n}$ and $\var{\tilde{s}}$ only if the determinant of the difference of the matrix and ${\rm \bf I}\,\omega^2/k^2$ vanishes, which sets a (complex) relation
\begin{align}
\omega^2 = c^2k^2-iDk^2\omega\,,\label{eq_w_of_c_def}
\end{align}
where the speed of sound $c$ and diffusivity $D$ are real quantities. Note that for decoupled modes [{\it e.g.} a pure density mode $(1,0)_{n,\tilde{s}}$] one recovers $c=\cc$ [{\it c.f.} Eq.~(\ref{eq_matrix_damping}) and Eq.~(\ref{eq_w_of_c_def})]. For mixed modes, however, both $c$ and $D$ are non-trivial functions of the transport coefficients. 

Both $\omega$ and $k$ are in general complex, and the phase velocity is given by $c_{\varphi}=\text{Re}(\omega)/\text{Re}(k)$, while $\gamma_x=\text{Im}(k)$ is the spatial attenuation constant, and $\gamma=-\text{Im}(\omega)$ is the temporal damping. The form of the waves is thus given by
\begin{align}
e^{i(kx-\omega t)} = e^{i[\text{Re}(k)x-\text{Re}(\omega) t]}e^{-\gamma_x x-\gamma t}\,,
\end{align}
and each variable has an amplitude and a phase ({\it e.g.} $\delta n=|\delta n|e^{i\varphi}$).

In Eq.~(\ref{eq_matrix_damping}), $\omega$ and $k$ appear independently, and not just as the ratio $\omega/k$, so it matters whether one solves for $\omega$ at fixed $k$ or {\it vice versa}. 

\subsection{Fixed-wavelength solutions}
\label{sect_fix_lambda}
We excite a standing wave with fixed (real) $k$, and observe a damped oscillation in time. In this case Eq.~(\ref{eq_w_of_c_def}) can be written as 
\begin{align}
\omega = \pm\sqrt{c^2k^2-\gamma^2}-i\gamma,\label{eq_complexw}
\end{align}
where we have inserted 
\vspace{-0.6em}
\begin{align}
\gamma=-\text{Im}(\omega)=\frac12Dk^2.
\end{align} 
We thus have 
\vspace{-0.6em}
\begin{align}
c&=|\omega|/k,\quad
c_{\varphi}=\text{Re}(\omega)/k
\end{align}
and, as for a damped harmonic oscillator, $\text{Re}(\omega)$ decreases with increasing $\gamma$ (at fixed $c$). 

To find $c$ and $\gamma$ in terms of the hydrodynamic parameters, we solve Eq.~(\ref{eq_matrix_damping}). Introducing dimensionless variables
\begin{align}
C \equiv \frac{\omega}{k\cc},\quad Q_{\kappa} \equiv \frac{\kappa}{c_V\cc}k,\quad Q_{\eta} \equiv \frac43\frac{\eta}{mn\cc}k,\label{eq_def_CQ_dimless}
\end{align}
and rewriting Eq.~(\ref{eq_matrix_damping}) as
\begin{align}
\pvec{
   -C^2 + 1  - i Q_{\eta}C & &  K^2/J\\
   -iJQ_{\kappa} & & -C-iQ_{\kappa}
}
\pvec{\frac{\delta n}{n} \\[8pt] \frac{\delta\tilde{s}}{\tilde{s}}} = 0,\label{eq_matrix_damping_CQ}
\end{align}
leads to a cubic polynomial in the normalized complex speed:
\begin{align}
C^3+iC^2(Q_{\kappa}+Q_{\eta})-C\braces{1+Q_{\kappa}Q_{\eta}}-iQ_{\kappa}(1-K^2)=0.\label{eq_charac_polynomial}
\end{align}
The solutions are invariant under the combined transformation $C\to-C$ and complex conjugation.
Two of the three roots are damped harmonic-oscillator-like solutions, which, depending on the magnitude of $Q_{\eta,\kappa}$, are underdamped, critically damped, or overdamped. The third root is always purely imaginary.
Most fluids are in the underdamped regime, where the pair of solutions [$C_{\rm I}^+=-(C^-_{\rm I})^*$] corresponds to right- and left- propagating pressure-density waves, and the third one ($C_{\rm II}$) is an overdamped entropy wave.

\begin{figure}[!t]
\centering
\includegraphics[width=0.99\columnwidth]{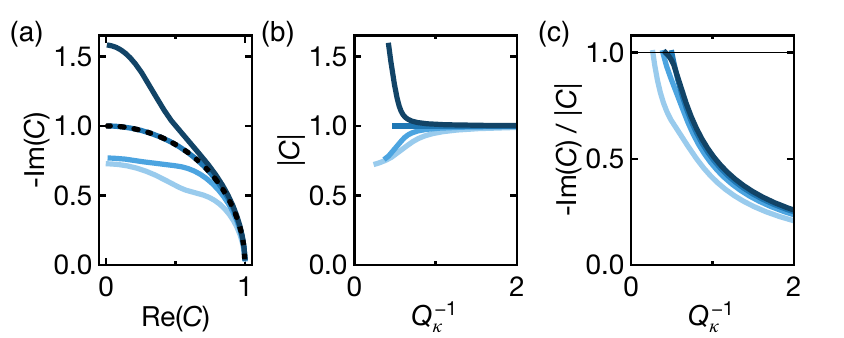}
\caption{
Damped speed of sound for $T>\Tc$ at fixed wavelength, showing the solution $C_{\rm I}^+$ of Eq.~(\ref{eq_charac_polynomial}) for varying $Q_{\kappa}$, at $K^2=1-c_V/c_P=0.5$ and $\text{Pr}=\{0.8,0.75,0.67,0.5\}$ (dark to light blue). (a) Parametric plot in the complex $C$ plane. The phase velocity, $\textrm{Re}(C_{\rm I}^+)$, decreases with increasing damping, $-\textrm{Im}(C_{\rm I})$. (b) The speed of sound $|C_{\rm I}|$ increases (decreases) for $\text{Pr}>3/4\, (\text{Pr}<3/4)$ with the strength of the transport coefficients. (c) Damping per cycle $-\textrm{Im}(C_{\rm I})/|C_{\rm I}|$, where a value of 1 corresponds to critical damping. The $\text{Pr}=0.67$ line of (b) and (c) is shown in Fig.~4 of the main text.
}
\label{fig_SIdamp}
\end{figure}

To connect to the regime of low damping, we expand the dimensionless solutions
in $Q_{\eta,\kappa}$ and use Eqs.~(\ref{eq_complexw},\,\ref{eq_def_CQ_dimless}) to find (up to second order in $\kappa$):
\begin{subequations}
\begin{align}
c_{\rm I} &= \cc|C_{\rm I}| = \cc - \frac{k^2}{2}\frac{K^2\kappa^2}{\cc c_Vc_P}\braces{1-\frac43\text{Pr}}\,,\\
\gamma_{\rm I} &= -k\cc \text{Im}(C_{\rm I}) = \frac{k^2}{2} \frac{\kappa}{c_P} \braces{\frac{4}{3}\text{Pr}-1+\frac{c_P}{c_V}}\label{eq_damping_lowQ}\,,\\
c_{\rm II}&=0,\quad \gamma_{\rm II} = -k\cc \text{Im}(C_{\rm II})= k^2\frac{\kappa}{c_P}\,.
\end{align}
\end{subequations}
Here we used $(1-K^2)=c_V/c_P$, and the Prandtl number 
\begin{align}
\label{eq_prandtl}
\text{Pr}\equiv \frac{c_P\eta}{mn\kappa} = \frac34\frac{Q_{\eta}c_P}{Q_{\kappa}c_V}
\end{align}
measures the relative weight of momentum diffusivity $\eta/(mn)$ to thermal diffusivity $\kappa/c_P$. 
The damping $\gamma_{\rm I}$, linear in the transport coefficients, is the usual Stokes--Kirchhoff damping, but the speed $c_{\rm I}$ is evaluated including a quadratic correction, which is relevant for $K\gg0$.
We see that $\text{Pr}^*=3/4$ determines if $c_{\rm I}$ decreases ($\text{Pr}<\text{Pr}^*$) or increases ($\text{Pr}>\text{Pr}^*$) with increasing damping. 
This result holds 
for the full solution of Eq.~(\ref{eq_charac_polynomial}) including all orders in $Q_{\eta,\kappa}$ because $C_{\rm I}^{+}=\sqrt{1-Q^2_{\kappa}/4}-iQ_{\kappa}/2$ is an exact solution at $\text{Pr}^*$, where $Q_{\eta}=Q_{\kappa}(1-K^2)$. 
Note that the phase velocity $c_{\varphi}=\cc\text{Re}(C_{\rm I}^+)$ always decreases with increasing damping for the case of waves with fixed wavelength.

For an ideal, classical gas, $\text{Pr}=2/3$, and for a weakly interacting Bose gas, Ref.~\cite{Nikuni_SI:1997} also predicts $\text{Pr}\approx2/3<\text{Pr}^*$, even close to $\Tc$. Qualitatively, $c_{\rm I}$ is therefore expected to decrease with increasing $\eta,\kappa$ (see Section~\ref{sect_fix_lambda}) consistent with our experiments [Fig.~4(b) in the main text]. 
One can also interpret our measurement of $c_{\rm I}$ as an indirect probe of $\text{Pr}$.
Assuming that the hydrodynamic theory holds down to $L/\mfp\approx 1.5$, we find $\text{Pr}=0.67(6)$ for our gas at $T/\Tc\approx1.3$.

Kinetic gas theory allows us to relate the damping to the mean free path~$\mfp$.
For a weakly interacting Bose gas, $\mfp=(8\pi na^2)^{-1}$ and the transport coefficients can be calculated from microscopics~\cite{Nikuni_SI:1997,Nikuni_SI:2001} based on the Chapman--Enskog method. For a fugacity $z$, to lowest order in $na\lambda_T^2$:
\begin{align}
Q_{\eta} &=\frac{\sqrt{15\pi}}{12}q_{\eta}(z)\,{k\mfp} \label{eq_Qeta_z},\\
Q_{\kappa} &=\frac{15}{64}\sqrt{15\pi}q_{\kappa}(z)\frac{n}{c_V}\,{k\mfp}.\label{eq_Qkappa_Pr}
\end{align}
Here $q_{\eta}(z)=\sqrt{g_{3/2}(z)/g_{5/2}(z)}\,\eta(z)/\eta(0)$ and $q_{\kappa}(z)=\sqrt{g_{3/2}(z)/g_{5/2}(z)}\,\kappa(z)/\kappa(0)$ capture the Bose effects (monotonic in $z$) and are defined such that $q_{\eta}(0)=\,q_{\kappa}(0)=1$.
Based on\cite{Nikuni_SI:1997}, one finds $q_\eta(1)=0.49$, $q_{\kappa}(1)=1.22$, while for $z<0.9$ one obtains $\text{Pr}\approx 2/3$ [via Eq.~(\ref{eq_prandtl})]. 

For weak damping, the damping per period from Eq.~(\ref{eq_damping_lowQ}), with Pr$=2/3$, is
\begin{align}
\frac{\gamma_{\rm I}}{c_{\rm I}k} = r k\mfp\,,\label{eq_damping_weak_Bose}
\end{align}
where
\vspace{-1em}
\begin{align}
r= \frac{\sqrt{15\pi}}{192}\left(9\frac{c_P}{c_V}-1\right)q_{\eta}(z)
\end{align}
is of order unity, so for $k\sim L^{-1}$ the damping per period is set by the Knudsen number $\mfp/L$.

It is instructive to express the diffusivity of a non-degenerate Bose gas in units of $\hbar/m$ [using $c_{\rm I}=\sqrt{2\pi \chi(z)}\,\hbar/(m\lambda_T)$, which is Eq.~(\ref{eq_cH_SI}) with $g=0$]:
\begin{align}
D_{\rm I}=2r \sqrt{2\pi \chi(z)}\frac{\mfp}{\lambda_T}\frac{\hbar}{m}\,. 
\end{align} 
The prefactor is again of order one, so $\mfp/\lambda_T$ sets the diffusivity in units of $\hbar/m$. Note that $\mfp$ is bounded from below by the interparticle distance $n^{-1/3}$, and a thermal gas always has $\lambda_T\lesssim n^{-1/3}$, so $D_{\rm I}\gtrsim\hbar/m$ for a gas above $\Tc$. 

In Fig.~4 of the main paper we show the full solution of Eq.~(\ref{eq_charac_polynomial}) in the underdamped regime 
using $z=0.75$, $c_V/c_P=0.5$, and $Q_{\eta,\kappa}$ from Eqs.~(\ref{eq_Qeta_z},\,\ref{eq_Qkappa_Pr}) with $q_{\eta}(0.75)=0.73$.
For this we obtain $r\approx0.44$, and for our smallest $\mfp\sim10\,\umu$m with $\lambda_T\sim1\,\umu$m a diffusivity $D\sim 30\,\hbar/m$.

\subsection{Fixed-frequency solutions}
\label{sect_fix_freq}
Finally, we briefly consider the opposite case of a fixed-frequency wave, {\it e.g.}~generated by a source oscillating at some (real) $\omega$; more details on this case can be found in~\cite{Temkin_SI:1981}. In this case, the wavelength forms dynamically and the damping occurs in space over a lengthscale $1/\gamma_x$. 

Using the same definition of $c$ and $D$ [Eq.~(\ref{eq_w_of_c_def})], one finds:
\begin{subequations}  
\begin{align}
k&=\frac{\omega}{\sqrt{c^2-iD\omega}},\\
c_{\varphi} &= 2\frac{\text{Re}(\sqrt{c^2+iD\omega})}{1+(1+D^2\omega^2/c^4)^{-1/2}},\\ 
\gamma_x &= \frac{\text{Im}(\sqrt{c^2+iD\omega})}{(c^4/\omega^2+D^2)^{1/2}}\,,
\end{align}
\end{subequations}
highlighting the main difference to the fixed-$k$ case: At constant $c$, the phase velocity $c_{\varphi}$ increases for increasing $D\omega$ (at fixed $\omega$), while it decreases for increasing $Dk^2$ (at fixed $k$).
This means that the sound wavelength increases with damping at fixed frequency, while the oscillation frequency decreases with damping at fixed wavelength. 

This illustrates the importance of carefully defining $c$ in the presence of  damping; solving Eq.~(\ref{eq_matrix_damping}), one can show that $c_{\rm I}$ and the damping ($\gamma_{\rm I}/c_{\rm cl},\ \gamma_{x,{\rm I}}$), are identical for fixed-wavelength and fixed-frequency scenarios up to quadratic order in $\kappa$ and $\eta$.

\clearpage
\section*{Appendix A: Landau--Khalatnikov equations}
\label{sec_lk_table}
The Landau--Khalatnikov hydrodynamic equations~\cite{Khalatnikov_SI:1965} (see also ~\cite{Griffin_SI:2009}) describe near-equilibrium excitations of any fluid. Within our notation (see Section~\ref{sec_si_theory}) the equations read 
\begin{subequations}
\begin{align}
\label{eq_landaukhalatnikov}
\partial_tn+\nabla\cdot\bm{j}&=0, \\
\partial_t\bm{v}_{\rm s}+\nabla\braces{\frac{\mu+V_{\rm ext}}{m}+\frac12v_{\rm s}^2}
&=M,\\
\partial_tj_{\mu} +\partial_{x_{\nu}}J_{\mu\nu}+\frac1{m}\braces{\partial_{x_{\mu}}p+n\partial_{x_{\mu}}V_{\rm ext}}&=G_{\mu}, \\
\partial_ts+\nabla\cdot\braces{s\bm{v}_{\rm n}-\frac{\kappa}{T}\nabla T} &= \frac{R}{T}\,,
\label{A3_EQ}
\end{align}
\end{subequations}
where sums over repeated indices are implied, 
\begin{align}
J_{\mu\nu} &= \ns v_{{\rm s}\mu}v_{{\rm s}\nu}+\nn v_{{\rm n}\mu}v_{{\rm n}\nu}\,,\\[5 pt]
D_{\mu\nu} &= \frac12\braces{\partial_{x_{\nu}}v_{{\rm n}\mu}+\partial_{x_{\mu}}v_{{\rm n}\nu}}\,,
\end{align}
and the source terms are
\begin{subequations}
\begin{align}
M&=\zeta_3m\nabla[\nabla\cdot\ns(\bm{v}_{\rm s}-\bm{v}_{\rm n})]+\zeta_4\nabla[\nabla\cdot\bm{v}_{\rm n}]\,,\\
G_{\mu} &= \partial_{x_{\nu}}\braces{2\frac{\eta}{m}[D_{\mu\nu}-\frac13\delta_{\mu\nu} (\nabla\cdot\bm{v}_{\rm n})]}\\\nonumber
&\phantom{=}+\partial_{x_{\mu}}\braces{\zeta_1\nabla\cdot[\ns(\bm{v}_{\rm s}-\bm{v}_{\rm n})]+\frac{\zeta_2}{m}\nabla\cdot \bm{v}_{\rm n}}\,,\\
R &= \zeta_2(\nabla\cdot\bm{v}_{\rm n})^2+2\zeta_1(\nabla\cdot\bm{v}_{\rm n})\nabla\cdot[m\ns(\bm{v}_{\rm s}-\bm{v}_{\rm n})]\nonumber\\
&\phantom{=} +\zeta_3\braces{\nabla\cdot[m\ns(\bm{v}_{\rm s}-\bm{v}_{\rm n})]}^2\\\nonumber
&\phantom{=}+2\eta\braces{D_{\mu\nu}-\frac13\delta_{\mu\nu} (\nabla\cdot\bm{v}_{\rm n})}^2 +\frac{\kappa}{T}(\nabla T)^2\,.\\\nonumber
\end{align}
\end{subequations}
The viscosities ($\eta,\zeta_{1,2,3,4}$) appear only in the source terms $M$, $G_{\mu}$, and $R$, whereas the heat conductivity $\kappa$ leads both to a transport of entropy (within the divergence term) and to the generation of entropy (within $R$).

Generally, several famous fluid equations are embedded in the Landau--Khalatnikov equations and can be recovered as limiting cases:
\begin{itemize}
\item {\bf Landau's two-fluid equations} are obtained when dropping dissipative processes ($\zeta_i=0,\kappa=0,\eta=0$).
\item Setting the superfluid density to zero ($\ns=0$) leads to the classical {\bf Navier--Stokes equations}. 
\item At zero temperature ({$T=0$, $\nn=0$, $\zeta_1 =\zeta_3 =0$}) one recovers the {\bf Gross--Pitaevskii equation} (in its hydrodynamic form).
\end{itemize}

Note that to describe our sound measurements we make two additional simplifications: (1) linearization, (2) treating the problem as effectively 1D, with all velocities aligned (which omits vortex dynamics).

\section*{Appendix B: Popov calculation of superfluid and normal densities}
\label{appendix:popov}
Here we outline how we solve the Popov equation for our system.
We first solve the Popov equation, $\mu=g(2n_T(\tilde{x})+n_{\rm BEC})$, for the dimensionless BEC density $\tilde{x}=gn_{\rm BEC}/(\kB T)$ using a grid of $\mu/(\kB T)$ values:
\begin{align}
\mu/(\kB T) &=  4\tilde{n}_T(\tilde{x}) a/\lambda_T + \tilde{x}\,.
\end{align}
Here, the dimensionless thermal density $\tilde{n}_T = n_T\lambda_T^3$ is
\begin{align}
\tilde{n}_T(\tilde{x}) &=\lambda_T^3\int (|u_k|^2+|v_{-k}|^2)f({\epsilon(k))\,{\rm d}\bm{k}^3},
\end{align}
where $f(\varepsilon)$ is the Bose function and $\epsilon(k)= \sqrt{\epsilon_0(k)^2 + 2gn_{\rm BEC}\epsilon_0(k)}$, with $ \epsilon_0(k)=\hbar^2k^2/(2m)$. Here $\epsilon(k)$, $u_k$, and $v_k$ are the Bogoliubov dispersion, particle- and hole-weights\cite{Pitaevskii_SI:2016}, which depend on $\tilde{x}$ via $n_{\rm BEC}$.

The densities $n_{\rm BEC}(\mu)$ and $n_T(\mu)$ are then integrated over the trap using the LDA:
\begin{align}
N_{\rm BEC} &= \int \mathcal{V}(\varepsilon) n_{\rm BEC}(\mu-\varepsilon)\,{\rm d}\varepsilon,\nonumber\\
N_T &= \int \mathcal{V}(\varepsilon) n_T(\mu-\varepsilon)\,{\rm d}\varepsilon,\nonumber\\
N&=N_T+N_{\rm BEC}\label{eqs_PopovN},
\end{align}
where $\mathcal{V}{\rm d}\varepsilon$ is the differential potential volume of our cylindrical power-law trap [Eq.~(\ref{eq_box_potential})]
\begin{align}
\mathcal{V}(\varepsilon) &= \frac{\partial}{\partial\varepsilon}\int_{V_{\rm box}<\varepsilon}{\rm d}\bm{r}^3
=\pi\,\chi_{\beta_r,\beta_z} R^2LV_r^{-\frac2{\beta_r}}V_z^{-\frac1{\beta_z}}\varepsilon^{\alpha-5/2}\nonumber,\\
\chi_{\beta_r,\beta_z} &= \frac{\Gamma(1+2/\beta_r)\Gamma(1+1/\beta_z)}{\Gamma(\alpha-3/2)},\quad
\alpha = \frac32+\frac2{\beta_r}+\frac1{\beta_z}.
\end{align}
For a given $N$, we then numerically solve Eq.~(\ref{eqs_PopovN}) for $\mu/(\kB T)$, which gives $n_{\rm BEC}(\bm{r})$ and $n_{T}(\bm{r})$ in the LDA.

Finally, we calculate normal and superfluid densities using Eqs.~(\ref{eq_popovnnns}).
\clearpage

\begin{widetext}
\section*{Appendix C: Thermodynamic quantities for a Bose gas in the Hartree--Fock approximation}
\label{sec_bose_table}
\vspace{-2.em}
\begin{table*}[h]
\caption{\normalsize 
Thermodynamic quantities for a homogeneous interacting Bose gas in the self-consistent Hartree--Fock mean-field approximation. The density $n$ is defined in a volume $V$, with $\nn=n_T$, $\ns= \nBEC$ and the excitation energy spectrum $\varepsilon_{\rm HF}(k)=\hbar^2 k^2/(2m)+2gn$, where $g=4\pi\hbar^2a/m$. The interaction-corrected chemical potential is $\mu^*=\mu-2gn$, which below $\Tc$ becomes $\mu^*=-g\nBEC$. For convenience we set $\kB=1$ and use the fugacity  $z=e^{\mu^*/T}$, the thermal wavelength $\lambda_T=\sqrt{2\pi\hbar^2/(m T)}$, and the polylogarithms $g_\alpha(z)$.
The approximation is valid for $gn\ll T$ since phonons are neglected and it requires $|\mu^*/T|\gtrsim (g/(\lambda_T^3 T))^2$ due to the breakdown of mean-field theory close to $\Tc$ (Ginzburg criterion).
  A higher accuracy can be reached by the Popov or Beliaev techniques\cite{Capogrosso-Sansone_SI:2010} and around $\Tc$ by Monte Carlo simulations\cite{Prokofev_SI:2004, Spada_SI:2021}.
  }
\vspace{0.5em}
{\fontsize{10 pt}{13 pt}\selectfont
\begin{tabular}{r|c|c|l|l}
Quantity &  & Definition 
&\multicolumn{1}{|c|}{Thermal Bose Gas} 
&\multicolumn{1}{|c}{BEC Phase}   \\
\hline
\rule{0pt}{3ex} 
Grand potential & $\Omega$ & $- T \ln(Z)$ 
& $-V\left(\frac{T}{\lambda_T^{3}}g_{5/2}(z)+gn^2\right)$
& $-V\left(\frac{T}{\lambda_T^{3}}g_{5/2}(z)+\frac12g(2n^2-\nBEC^2)\right)$
\\
\rule{0pt}{3ex} 
Free Energy  & $F$ & $\Omega+\mu N$ 
& $-V\left(\frac{T}{\lambda_T^{3}}g_{5/2}(z)-gn^2-nT\ln(z)\right)$
& $-V\left(\frac{T}{\lambda_T^{3}}g_{5/2}(z)-\frac12g\left(n^2+n_T^2\right)\right)$\\
\rule{0pt}{3ex} 
Energy & $E$ & $F+T S$ 
& $\phantom{-}V\left(\frac32nT\frac{g_{5/2}(z)}{g_{3/2}(z)}+gn^2\right)$
& $\phantom{-}V\left(\frac32n_TT\frac{g_{5/2}(z)}{g_{3/2}(z)}+\frac12g(2n^2-\nBEC^2)\right)$\\
\rule{0pt}{3ex} 
\begin{minipage}[c]{2cm} \flushright{Chemical potential}\end{minipage}
&$\mu$& $\frac1{V}\drvTD{F}{n}{T,V} $ 
&  $\phantom{-}2gn+T\ln(z)$
&  $\phantom{-}2gn-g\nBEC $\\
\rule{0pt}{3ex} 
Density &$n$& $-\frac1{V}\drvTD{\Omega}{\mu}{T,V} $ 
&  $\phantom{-}\frac1{\lambda_T^3}g_{3/2}(z)$
&  $\phantom{-}n_T+\nBEC = \frac1{\lambda_T^3}g_{3/2}(z)-\frac{T}{g}\ln(z)$\\
\rule{0pt}{3ex} 
\begin{minipage}[c]{2cm} \flushright{Phase space density}\end{minipage} 
&$\varpi$& $n\lambda_T^3$ 
&  $\phantom{-}g_{3/2}(z)$
&  $\phantom{-}g_{3/2}(z)-\frac12\frac{\lambda_T}{a}\ln(z)$\\
\rule{0pt}{3ex} 
Entropy &$S$& $-\drvTD{F}{T}{N,V} $ 
&  $\phantom{-}V\left(\frac52\frac1{\lambda_T^3}g_{5/2}(z)-n \ln(z)\right)$
&  $\phantom{-}V\left(\frac52\frac1{\lambda_T^3}g_{5/2}(z)-n_T \ln(z)\right)$\\
\rule{0pt}{3ex} 
\begin{minipage}[c]{2cm} \flushright{Entropy per particle}\end{minipage} 
&$\tilde{s}$& $\frac{S}{nV} $
&  $\phantom{-}\frac52\frac{g_{5/2}(z)}{g_{3/2}(z)}- \ln(z)$
&  $\phantom{-}\frac{n_T}{n}\left(\frac52\frac{g_{5/2}(z)}{g_{3/2}(z)}- \ln(z)\right)$
\\
\rule{0pt}{3ex} 
Pressure &$p$& $-\frac{\Omega}{V}$ 
& $\phantom{-}\frac{T}{\lambda_T^3}g_{5/2}(z)+gn^2$
& $\phantom{-}\frac{T}{\lambda_T^{3}}g_{5/2}(z)+gn^2-\frac12g\nBEC^2$
\\
\rule{0pt}{3ex} 
Heat capacity &$c_V$& $\frac{T}{V}\drvTD{S}{T}{V,N}$ 
& $\phantom{-}n\left(\frac{15}{4}\frac{g_{5/2}(z)}{g_{3/2}(z)}-\frac94\frac{g_{3/2}(z)}{g_{1/2}(z)}\right)$
& $\phantom{-}n_T\left(\frac{15}{4} \frac{g_{5/2}(z)}{g_{3/2}(z)}+3\frac{g}{T}\left(n-\frac14n_T\right)\right)+\mathcal{O}(g^{\tfrac32})$
\\
\rule{0pt}{3ex} 
Heat capacity &$c_P$& $\frac{T}{V}\drvTD{S}{T}{p,N}$ 
& $\phantom{-}c_V\frac{\kappa_T}{\kappa_S}$
& $\phantom{-}\frac{25}{4}\frac{g^2_{5/2}(z)}{\varpi^2\lambda_T^3}\braces{\frac{T\lambda_T^3}{g}+g_{1/2}(z)}+\mathcal{O}(g^{0})$
\\
\rule{0pt}{3ex} 
Compressibility &$\kappa_S$& $\frac1{n}\drvTD{n}{p}{S,N}$ 
& $\phantom{-}\frac1{n}\left(\frac53\frac{g_{5/2}(z)}{g_{3/2}(z)}T+2gn\right)^{-1}$
& $\phantom{-}\left(\frac53\frac{T}{\lambda_T^3}g_{5/2}(z)+g(2n^2\hspace{-0.5mm}-\hspace{-0.5mm}\nBEC^2)\right)^{{-}1}\hspace{-1.5mm}+\mathcal{O}(g^{\frac32})$
\\
\rule{0pt}{3ex} 
Compressibility &$\kappa_T$& $\frac1{n}\drvTD{n}{p}{T,N}$ 
& $\phantom{-}\hspace{-0.5mm}\frac1{n}\left(\frac{g_{3/2}(z)}{g_{1/2}(z)}T+2gn\right)^{-1}$
& $\phantom{-}\frac1{gn^2}\frac{T-g\lambda_T^{-3} g_{1/2}(z)}{T-2g\lambda_T^{-3} g_{1/2}(z)}$
\\
\rule{0pt}{3ex} 
\begin{minipage}[c]{2cm} \flushright{Thermal expansion}\end{minipage}
&$\alpha_T$& $-\frac1{n}\drvTD{n}{T}{p,N}$ 
& $\phantom{-}\frac{5g_{1/2}(z)g_{5/2}(z)-3g_{3/2}^2(z)}{4gn\ g_{1/2}(z)g_{3/2}(z)+2Tg_{3/2}^2(z)}$
& $\phantom{-}\frac{5}{2}\frac{g_{5/2}(z)}{\varpi^2}\left(\frac{\lambda_T^3}{g}+\frac1{T}g_{1/2}(z)\right) +\mathcal{O}(g^{0})$
\\
\rule{0pt}{4ex} 
\begin{minipage}[c]{2cm} \flushright{Heat capacity ratio}\end{minipage}
&$\gamma_c$& $\frac{c_P}{c_V},\frac{\kappa_T}{\kappa_S}$ 
& $\phantom{-}\frac{5}{3\varpi^2}\,\frac{\lambda_Tg_{5/2}(z)}{\lambda_Tg^{-1}_{1/2}(z)+4a} + \frac{4a}{\lambda_Tg^{-1}_{1/2}(z)+4a}$
& $\phantom{-}\frac53\frac{g_{5/2(z)}}{\varpi^2}\braces{\frac{T\lambda_T^3}{g}+g_{1/2}(z)}+\mathcal{O}(g^{0})$
\\
\rule{0pt}{4ex} 
\begin{minipage}[c]{2cm} \flushright{}\end{minipage}
&$K$& $\sqrt{1{-}\frac1{\gamma_c}},\sqrt{\frac{\alpha_T^2T}{\kappa_Tc_P}}$ 
& $\phantom{-}\frac{5g_{5/2}(z)-3g^2_{3/2}(z)g_{1/2}^{-1}(z)}{5g_{5/2}(z)+6\tfrac{gn}{T}g_{3/2}(z)}$
& $ \phantom{-} 1-\frac{3}{10}\frac{g_{3/2}(z)}{g_{5/2}(z)}\frac{gn^2}{n_T T}+\mathcal{O}(g^{3/2})$
\\
\rule{0pt}{4ex} 
\begin{minipage}[c]{2cm} \flushright{}\end{minipage}
&$J$& $- \frac{n}{\tilde{s}}\drvTD{\tilde{s}}{n}{T,N}$ 
& $\phantom{-}\frac{5g_{5/2}(z)-3g^2_{3/2}(z)g_{1/2}^{-1}(z)}{5g_{5/2}(z)\,-2g_{3/2}(z)\ln(z)}$
& $ \phantom{-}1+\frac{3}{5}\frac{g_{3/2}(z)}{g_{5/2}(z)}\frac{gn}{T}+\mathcal{O}(g^{2})$
\\
\rule{0pt}{4ex} 
\begin{minipage}[c]{2.5cm} \flushright{Classical speed of sound}\end{minipage}
&$\cc$& $\sqrt{\frac1{mn\kappa_S}}$ 
& $\phantom{-}\sqrt{\frac53\frac{g_{5/2}(z)}{g_{3/2}(z)}\frac{T}{m}+2\frac{gn}{m}}$
& $\sqrt{\frac{5}{3}\frac{g_{5/2}(z)}{g_{3/2}(z)}\frac{n_TT}{mn} + \frac{g}{m}\left(2n-\frac{\nBEC^2}{n}\right)} +\mathcal{O}(g^{3/2})$

\\
\rule{0pt}{4ex} 
\begin{minipage}[c]{2cm} \flushright{}\end{minipage}
&$b$& $\cc\frac{K}{J},\,\sqrt{\frac{Ts^2}{c_Vn}}$ 
& $\phantom{-}\frac{{5g_{5/2}(z)-2\varpi\ln(z)}}{\sqrt{5 g_{5/2}(z)-3\varpi^2g_{1/2}^{-1}(z)}}\sqrt{\frac{T}{3\varpi}}$
& $\sqrt{\frac{5}{3}\frac{g_{5/2}(z)}{g_{3/2}(z)}\frac{n_TT}{mn}-\frac{gn_T^2}{mn}}+\mathcal{O}(g^{2})$
\\
\rule{0pt}{4ex} 
\begin{minipage}[c]{1.6cm} \flushright{Speed of first sound}\end{minipage}
&$c_{\rm I}$& 
 \begin{minipage}[c]{2cm} \flushright{see Eq.~(\ref{eq_generalFirstSound}) }\end{minipage}
& $\phantom{-}\sqrt{\frac53\frac{g_{5/2}(z)}{g_{3/2}(z)}\frac{T}{m}+2\frac{gn}{m}}$
& $\sqrt{\frac{5}{3}\frac{g_{5/2}(z)}{g_{3/2}(z)}\frac{T}{m} + 2\frac{g\nn}{m}} +\mathcal{O}(g^{3/2})$
\\
\rule{0pt}{4ex} 
\begin{minipage}[c]{2cm} \flushright{Speed of second sound}\end{minipage}
&$c_{\rm II}$& 
 \begin{minipage}[c]{2cm} \flushright{see Eq.~(\ref{eq_generalSecondSound}) }\end{minipage}
& $\phantom{-} 0$
& $\phantom{-}\sqrt{\frac{g\ns}{m}}+\mathcal{O}(g^1)$
\end{tabular}
}
\label{table_HF}
\end{table*}

\newpage
\clearpage

\end{widetext}

%


\end{document}